\documentclass[aps,pra,reprint,groupedaddress,superscriptaddress,amsmath,amssymb]{revtex4-2}
\usepackage{graphicx}
\usepackage{bm}
\usepackage{color}
\usepackage[colorlinks=true,linkcolor=blue,citecolor=blue,urlcolor=blue]{hyperref}

\graphicspath{{./}{../../figures/}{../figures/}}

\newcommand{\ad}{\hat a^{\dagger}}
\newcommand{\bd}{\hat b^{\dagger}}
\newcommand{\jp}{\hat J_{+}}
\newcommand{\jm}{\hat J_{-}}
\newcommand{\jz}{\hat J_{z}}
\newcommand{\kp}{\hat K_{+}}
\newcommand{\km}{\hat K_{-}}
\newcommand{\kz}{\hat K_{z}}
\newcommand{\Ne}{\hat N_{\mathrm{e}}}
\newcommand{\nb}{\bar n_0}

\newcommand{\Tr}{\mathrm{Tr}}
\newcommand{\ket}[1]{\left| #1 \right\rangle}
\newcommand{\bra}[1]{\left\langle #1 \right|}

\newcommand{\mean}[1]{\left\langle #1 \right\rangle}

\begin{document}

\title{Charging Dicke and Tavis--Cummings quantum batteries with a
$\mathcal{PT}$-symmetric lossy--gain cavity pair}

\author{Shiqing Tang}
\email{sqtang@hynu.edu.cn}
\affiliation{College of Physics and Electronic Engineering, Hengyang Normal University, Hengyang 421002, China}
\author{Qing Yu}
\affiliation{School of Information Science and Engineering, Hunan Institute of Engineering, Xiangtan 411104, China}
\author{Ying Li}
\affiliation{School of Information Science and Engineering, Hunan Institute of Engineering, Xiangtan 411104, China}
\author{Cuilu Zhai}
\affiliation{Department of Maths and Physics, Hunan Institute of Engineering, Xiangtan 411104, China}
\author{Zhao-Hui Peng}
\affiliation{Hunan Provincial Key Laboratory of Intelligent Sensors and Advanced Sensor Materials, and Department of Physics, Hunan University of Science and Technology, Xiangtan 411201, China}
\author{Wangjun Lu}
\email{wjlu1227@zju.edu.cn}
\affiliation{School of Information Science and Engineering, Hunan Institute of Engineering, Xiangtan 411104, China}

\date{\today}

\begin{abstract}
We study how a parity--time-symmetric photonic environment changes the
charging of Dicke and Tavis--Cummings quantum batteries.  For atoms
prepared in the collective ground state the excitation number grows as
the square of the charging time, with a coefficient set by the initial
mean photon number; the counter-rotating terms of the Dicke model add
one to four times that number.  Conservation of the total excitation
number turns this into an exact ceiling on the stored Tavis--Cummings
energy, the transition energy times the smaller of the photon and atom
numbers, met numerically to within two per cent, which the Dicke terms
overshoot by a few per cent at ultrastrong coupling.  We then couple a
battery to a lossy--gain pair, keeping the parity--time symmetry intact
by putting an identical battery in each cavity; the closed moment
system puts the exceptional point at the loss rate.  Below it the gain
cavity fills and the gain-side battery charges while the lossy-side one
stays empty; above it photons are shuttled away and dissipated.
Solving the full Lindblad equation with batched Monte--Carlo wave
functions, we find that the gain-side battery gains thirty-nine,
twenty-six and twenty-eight per cent in stored energy, power and
ergotropy in the broken phase and loses sixty-five, thirty-five and
sixty-five per cent deep in the unbroken phase, where the lossy-side
battery overtakes it.  Restoring the counter-rotating terms leaves all
of this unchanged within the statistical error; what they add is a
reactive oscillation at twice the transition frequency that does no net
work but nearly doubles the peak of the instantaneous power.  The extra
energy is of lower quality, the recoverable fraction dropping from
about ninety-eight to ninety per cent.  The optimum sits on the broken
side, close to but not at the exceptional point.
\end{abstract}

\maketitle

\section{Introduction}\label{sec:intro}

A quantum battery is a finite quantum system that stores energy supplied
by an external charger and releases a part of that energy as
work~\cite{Alicki2013,Binder2015,Campaioli2024}.  The scaling of the
charging power with the number of cells is the central issue of the
field, and it was settled early on by the observation that letting many cells share
a single quantum charger produces a collective, rather than merely
additive, speed-up~\cite{Ferraro2018,Campaioli2017}.  This quantum
charging advantage has since been bounded from first
principles~\cite{JuliaFarre2020} and shown to require entangling operations
that act globally on the cells~\cite{Gyhm2022}; it has been related to the
entanglement~\cite{Gyhm2024} and to the coherence~\cite{Francica2020} of
the state produced during charging.  Demonstrations cover a range of
many-body platforms: spin chains~\cite{Le2018}, Sachdev--Ye--Kitaev and
other fast-scrambling models~\cite{Rossini2020,Rosa2020}, disordered and
localized systems~\cite{Rossini2019}, harmonically and periodically
driven setups~\cite{Zhang2019}, adiabatic protocols designed to be stable
against disorder~\cite{Cakmak2019}, and hyperbolic
lattices~\cite{Downing2024}.  Experimentally the collective regime has
been reached in an organic microcavity showing
superabsorption~\cite{Quach2022} and in related organic
architectures~\cite{Li2025}.  Dicke superradiance rests on the same
cooperative-radiation physics~\cite{Dicke1954,Scully2009}.  The states
from which no work can be extracted at all are classified by the
thermodynamic theory of passive
states~\cite{Lenard1978,Pusz1978,Allahverdyan2004}.

Much of the literature treats the charger and the battery jointly as a
closed bipartite system.  Both parts of the problem are then
unitary: the energy delivered by the charger is bounded, the useful work
is governed by the passivity properties of the final
state~\cite{Perarnau2015,Alimuddin2020}, and no further energy enters
after the charging begins.

Batteries are, however, inevitably open.  Two very different
open-system effects then compete.  Decoherence degrades the extractable
work~\cite{Andolina2019}: even a weak coupling to the environment during
the charging converts coherent collective dynamics into an incoherent
cascade and eats into the collective advantage.  A driven
environment does the opposite and pumps energy into the battery.
Charging of that kind has been studied as charger-mediated energy
transfer~\cite{Farina2019,Andolina2019b} and as a protocol in which the
environment is the resource rather than the enemy~\cite{Barra2019,Mayo2022},
as well as in non-Markovian settings where memory effects slow down
self-discharging~\cite{Tabesh2020,Kamin2020}, as coherence storage in open
batteries~\cite{Caravelli2021}, and for the ubiquitous single-atom loss
and dephasing of Dicke and Tavis--Cummings batteries~\cite{Canzio2025}.
A collective enhancement of the dissipative charging itself has been
reported too~\cite{Pokhrel2025}, and the whole line of work sits
inside the broader framework of quantum thermal
machines~\cite{Mitchison2019,Bhattacharjee2021}.  The balance between
the two effects is set by the structure of the environment, not only by
its strength.

We address it with a setting in which the structure is both sharp and
tunable: a $\mathcal{PT}$-symmetric pair of cavities.
The theory of non-Hermitian degeneracies goes back to the observation
that certain non-Hermitian Hamiltonians nevertheless possess entirely
real spectra~\cite{Bender1998,Bender1999,Bender2007}, which was later
understood as a consequence of an antilinear rather than a Hermitian
symmetry~\cite{Mostafazadeh2002}.  At an exceptional point (EP) two
eigenvalues and their eigenvectors coalesce, a phenomenon studied
in detail in model systems~\cite{Heiss2000,Heiss2012}, seen directly in
microwave cavities~\cite{Dembowski2001} and in coupled
waveguides~\cite{Klaiman2008}, and reviewed from the open-systems
standpoint in Refs.~\cite{Rotter2009,Ashida2020,Bergholtz2021}.  Photonics
has been the main experimental arena for this physics: passive
$\mathcal{PT}$ symmetry was realised with balanced gain and loss in
coupled waveguides~\cite{Guo2009,Ruter2010} and subsequently in
active--passive coupled microresonators~\cite{Chang2014,Peng2014}, giving
access to loss-induced suppression and revival of
lasing~\cite{Peng2014b}, $\mathcal{PT}$-symmetric microring
lasers~\cite{Hodaei2014}, pump-induced exceptional points in
lasers~\cite{Liertzer2012,Brandstetter2014}, and EP-enhanced
sensing~\cite{Wiersig2014,Hodaei2017}; reviews are
available~\cite{ElGanainy2018,Ozdemir2019,Feng2017,Longhi2018,Miri2019,Ding2022}.

In cavity quantum electrodynamics the situation is subtler, because gain
and loss are genuine Liouvillian channels rather than a modification of
the Hamiltonian.  When the quantum fluctuations of a driven-dissipative
nonlinear oscillator are eliminated, the resulting semiclassical
description does display $\mathcal{PT}$ symmetry and an associated
driven-dissipative phase transition~\cite{Zhang2021}; the quantum-noise
price of an actual linear amplifier, on the other hand, has been
understood since the early work on amplifier
noise~\cite{Caves1982,Clerk2010}.  Balancing gain against loss, which is
reservoir engineering rather than Hamiltonian non-Hermiticity, is also
the standard route to directional, nonreciprocal photon
transport~\cite{Metelmann2015}, and there are now quantitative statements
about what non-Hermitian amplification can and cannot buy once quantum
noise is accounted for~\cite{Lau2018}.

Non-Hermitian charging protocols themselves have only begun to be
explored.  They have been studied mostly for lattice batteries, where
topology and non-Hermiticity compete~\cite{Zhou2026}, and for wireless
energy transfer through a non-Hermitian medium~\cite{Yang2025}.  The
cavity-QED version studied here differs from both.  The non-Hermiticity
is carried entirely by the charger, so the battery Hamiltonian
stays Hermitian, and the change of spectral regime happens at an
exceptional point of the second-moment dynamics, which we obtain
exactly.  Since the realisation of passive $\mathcal{PT}$ symmetry in
coupled optical resonators, it has been known that a balanced
loss--gain pair has
two qualitatively different regimes separated by that point: when the
inter-cavity coupling exceeds the loss/gain rate the relevant
eigenvalues are purely imaginary (unbroken phase), while in the
opposite case they acquire a real part (broken phase).  In the broken phase
the two modes can no longer exchange population efficiently, and the gain
mode accumulates energy.

We place a Dicke or TC quantum battery in each cavity of such a pair, the
two batteries being identical so that the $\mathcal{PT}$ symmetry of the
generator is kept intact, and compare the charging of the two batteries
on either side of the EP.  Each battery is charged by
whatever photons are in its cavity: in the broken phase the gain cavity
keeps them and keeps collecting more, so the gain-side battery charges
and the lossy-side one stays empty, while in the unbroken phase the
strong coherent coupling shuttles the photons into the lossy cavity,
where the lossy-side battery intercepts them on their way to the drain.
The numbers below bear this out.  The
unbroken phase carries a second penalty as well, and an independent one:
hybridising the two cavities detunes the normal modes from the atomic
transition.

The paper is organised as follows.
Section~\ref{sec:part1} deals with the isolated Dicke and TC batteries
charged by an initial photonic reservoir: it derives the short-time law
for the number of excited atoms, together with an exact bound on the
stored energy, and shows that the numerics recover the short-time
coefficient while the Dicke model violates the bound at ultrastrong
coupling.  Section~\ref{sec:part2} takes the cavity pair on its own and
gives its exact moment system, hence the position of the exceptional
point, the growth rate in the broken phase and the secular drift of the
unbroken phase; the generator is reproduced by an independent
construction.  Section~\ref{sec:twobat} couples a battery to the pair
without breaking its $\mathcal{PT}$ symmetry, by putting an identical
$N$-atom battery in each cavity, and shows that the broken phase still
delivers its energy to the gain-side battery alone, while in the
unbroken phase the lossy-side one overtakes it; the rest of that
section compares the two batteries in detail, scanning the
inter-cavity coupling across the exceptional point for the peak stored
energy, peak power and ergotropy, the time traces behind them, the
mechanism that decides which of the two batteries receives the energy,
and how much of either charge can be recovered as work.
Section~\ref{sec:dicke} repeats the whole comparison with the Dicke
coupling, that is, with the counter-rotating terms that
Eq.~\eqref{eq:symH} omits, and finds that at the coupling used
throughout they move every observable by less than the statistical
error; that is what licenses the Tavis--Cummings treatment of
Sec.~\ref{sec:twobat}, and it is why the two models can be discussed
together from the outset.
Section~\ref{sec:ep} then takes up the exceptional point itself and
resolves it in closed form in the low-excitation
limit: the Tavis--Cummings batteries leave it at $J=\kappa$ for every
coupling, halving the growth rate and moving the critical eigenvalues
to $\pm i\lambda$, the Dicke batteries shift it by an amount of order
$(\lambda/2\omega_0)^2$, and unequal loading destroys the symmetric
phase altogether.
Section~\ref{sec:method} then describes the batched Monte-Carlo
wave-function engine used throughout.  Section~\ref{sec:concl}
summarises, and three appendices
give the short-time expansion (App.~\ref{app:shorttime}), the moment
system (App.~\ref{app:moments}) and the convergence study
(App.~\ref{app:numeric}).

\section{The initial photonic reservoir}\label{sec:part1}

\subsection{Models}\label{sec:models}

The cavity mode $\hat a$ has frequency $\omega$ and the atoms
$\omega_0$; they are on resonance, $\omega=\omega_0$, and we use units
with $\omega_0=1$.  For $N$ identical two-level atoms restricted
to the symmetric Dicke manifold~\cite{Dicke1954,Tavis1968,Kirton2019,Garraway2011}
$j=N/2$, the collective spin operators
$\jz,\jp,\jm$ obey $[\jz,\jp]=\jp$ and $\jp\ket{j,m}=\sqrt{(j-m)(j+m+1)}
\ket{j,m+1}$.  The two battery Hamiltonians are
\begin{align}
\hat H_{\rm TC} &= \omega\ad\hat a+\omega_0\jz + \frac{\lambda}{\sqrt N}
\left(\hat a\jp+\ad\jm\right),\label{eq:htc}\\
\hat H_{\rm Dic} &= \omega\ad\hat a+\omega_0\jz + \frac{\lambda}{\sqrt N}
\left(\hat a+\ad\right)\left(\jp+\jm\right).\label{eq:hdic}
\end{align}
Both keep the field free energy $\omega\ad\hat a$.  Removing it is not
an approximation one may or may not make; it is the change of frame
generated by $U(t)=\exp[i\omega t(\ad\hat a+\jz)]$, and what that
change costs is decided by whether the coupling conserves the
excitation number.  For Eq.~\eqref{eq:htc} it does,
$[\jz,\hat H_{\rm TC}]=0$, so at resonance the free terms generate
nothing but a phase, the laboratory and rotating frames give identical
answers for every observable below, and we drop them from
Sec.~\ref{sec:part2} onwards.  For Eq.~\eqref{eq:hdic} it does not:
$[\jz,\hat H_{\rm Dic}]\neq0$, and removing the free energies promotes
$\hat a\jm$ and $\ad\jp$ from being off resonant by $2\omega_0$ to
being resonant, which changes the model rather than the frame.
Equation~\eqref{eq:hdic} is therefore integrated as written in this
section, and its rotating-frame image, used in
Sec.~\ref{sec:dicke}, carries the explicit factors $e^{\pm2i\omega_0t}$
of Eq.~\eqref{eq:dickeH}.
Throughout we write $\Ne\equiv\jz+j$ for the number of excited atoms, so
that the stored energy is $E(t)=\omega_0\mean{\Ne(t)}$ and the state of
all atoms in the ground level is $\Ne=0$.  The reservoir is either a
coherent state $\ket{\alpha}$ with $\bar n_0=|\alpha|^2$ or a Fock
state $\ket{n}$; throughout we take $N=10$ and $\lambda=0.1\,\omega_0$,
with $\lambda=0.3\,\omega_0$ for the ultrastrong-coupling comparison of
Sec.~\ref{sec:bound}.  Equation~\eqref{eq:hdic} retains both rotating and
counter-rotating terms; in the ultrastrong-coupling regime where those
terms matter it is the same model that underlies the equilibrium and
nonequilibrium superradiant transition~\cite{Hepp1973,Emary2003}, whose
driven and parametrically driven versions show rich dissolution and
dynamical phases~\cite{Bhaseen2012,Chitra2015}.

\begin{figure*}[tbp]
\centering
\includegraphics[width=\textwidth]{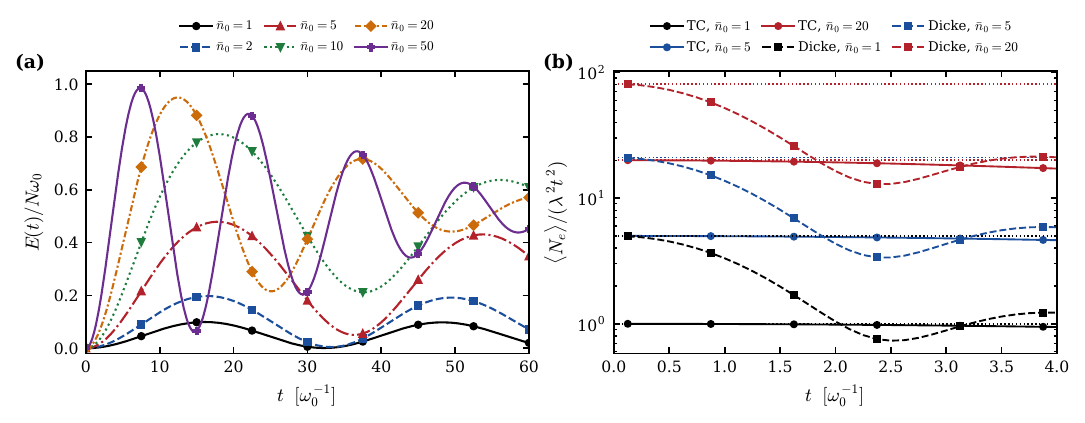}
\caption{(Color online) Isolated batteries, $N=10$ atoms,
$\lambda=0.1\,\omega_0$, all atoms initially in $\ket{j,-j}$.
(a) Stored energy $E(t)/N\omega_0$ of the TC model for six
coherent reservoirs.  Linestyles and markers (identical in all figures):
$\nb=1$ black solid circles, $\nb=2$ blue dashed squares,
$\nb=5$ red dash-dotted up-triangles, $\nb=10$ green dotted
down-triangles, $\nb=20$ orange short-dash diamonds, $\nb=50$ purple
solid squares.  The peak increases monotonically with $\nb$ and
saturates at full inversion.
(b) Short-time scaling test.  The ordinate is
$\mean{\Ne(t)}/(\lambda^2t^2)$; at $t\to0$ it must approach $\bar n_0$
for the TC model (solid, circles) and $4\bar n_0+1$ for the Dicke model
(dashed, squares), shown for $\nb=1$ (black), $5$ (blue) and $20$
(red).  Horizontal dotted lines mark the theoretical values of
Eqs.~\eqref{eq:shortTC} and \eqref{eq:shortDicke}.  Every curve starts
on its dotted line and peels away later, the departure time being set by
$\lambda\sqrt{4\nb+1}\,t\sim1$, i.e. by the depletion of the
reservoir.}
\label{fig:charging}
\end{figure*}

\subsection{Short-time perturbation theory}\label{sec:shorttime}

Let $\ket{\psi_0}=\ket{\mathrm{field}}\otimes\ket{j,-j}$, so that
$\Ne\ket{\psi_0}=0$.  Expanding $\Ne(t)$ to second order in $t$
(App.~\ref{app:shorttime}) gives
\begin{equation}
\mean{\Ne(t)}=t^2\bra{\psi_0}\hat H\Ne \hat H\ket{\psi_0}+O(t^4),
\label{eq:shortgeneral}
\end{equation}
because the terms linear in $t$ and the $\bra{\psi_0}\hat H^2\Ne\ket{\psi_0}$
term both vanish identically.  Since
$\hat H_{\rm TC}\ket{\psi_0}=(\lambda/\sqrt N)\sqrt N\,\hat a\ket{\mathrm{field}}
\otimes\ket{j,-j+1}=\lambda\,\hat a\ket{\mathrm{field}}\otimes\ket{j,-j+1}$,
and $\Ne$ acts as the unit operator on $\ket{j,-j+1}$,
Eq.~\eqref{eq:shortgeneral} yields
\begin{equation}
\mean{\Ne(t)}_{\rm TC}=\lambda^2\bar n_0\,t^2+O(t^4)
\label{eq:shortTC}
\end{equation}
for a coherent reservoir (and $\lambda^2 n\,t^2$ for a Fock reservoir).
For the Dicke model the same step gives
$\mean{\Ne(t)}_{\rm Dic}=\lambda^2
\mean{(\hat a+\ad)^2}_0\,t^2+O(t^4)$, i.e.
\begin{equation}
\mean{\Ne(t)}_{\rm Dic}=\lambda^2(4\bar n_0+1)\,t^2+O(t^4)
\label{eq:shortDicke}
\end{equation}
for a coherent reservoir and $\lambda^2(2n+1)t^2$ for a Fock reservoir.
Both coefficients increase monotonically with $\bar n_0$; this is the
perturbative version of the statement that a larger initial photon
number excites more atoms.  The factor $4$ in Eq.~\eqref{eq:shortDicke} is the
counter-rotating contribution $\mean{\hat a^2+\ad{}^2}_0=2\bar n_0$ (for real
$\alpha$) added to the rotating contribution $2\bar n_0$.

\subsection{The excitation-number bound}\label{sec:bound}

For the TC model the total excitation number
$\hat M=\ad \hat a+\jz$ is conserved, $[\hat H_{\rm TC},\hat M]=0$.  Hence
\begin{equation}
\mean{\Ne(t)}=\nb-\mean{\ad \hat a}_t\;\le\;\nb,
\qquad
\mean{\Ne(t)}\;\le\;N,
\label{eq:bound}
\end{equation}
so that $E_{\max}\le\omega_0\min(\nb,N)$ exactly, for every
initial field state and every time.  Full inversion of the battery
therefore requires at least $N$ photons in the reservoir.  This bound is
about the reservoir rather than the coupling, and it is the reason why
``more initial photons'' yields more charge and not merely faster
charging.  Two caveats attach to this bound.  It constrains the resources
initially present and not any particular protocol:
how best to distribute a fixed photon budget is a separate optimisation
problem, solved for the TC battery by Bethe-ansatz techniques in
Ref.~\cite{Lu2021}, while bounds on capacity and power that hold
independently of the protocol were derived in Ref.~\cite{JuliaFarre2020}.
It is also a statement about a
closed system only.  Nothing in Eq.~\eqref{eq:bound} survives
contact with an environment that exchanges excitations with the battery,
and arranging just such an environment is the subject of
Sec.~\ref{sec:part2}.

The Dicke model has no such conservation law, because the
counter-rotating piece $\hat a\jm+\ad\jp$ creates and annihilates photons in
pairs with the spin.  Equation~\eqref{eq:bound} can then be violated, a
point we return to below; violations of the same origin have been
reported for two-photon Dicke
charging~\cite{Crescente2020} and for charging an off-resonant Dicke
battery by virtual photons~\cite{Gemme2023}.

\subsection{Semiclassical limit}\label{sec:semi}

For $\nb\gtrsim N$ the reservoir is barely depleted and may be replaced,
in the sense of the correspondence between a quantised mode and a
$c$-number field~\cite{Jaynes1963,ScullyZubairy}, by a classical drive
of amplitude $\sqrt{\nb}$, which reduces
Eq.~\eqref{eq:htc} to a collective rotation about $\hat J_x$ at
$\Omega=2\lambda\sqrt{\nb/N}$.  Then
\begin{align}
\mean{\Ne(t)}&=\frac{N}{2}\left[1-\cos\left(\Omega t\right)\right],&
t_{\rm inv}&=\frac{\pi\sqrt N}{2\lambda\sqrt{\nb}},
\label{eq:semi}\\
P_{\max}&=\omega_0\lambda\sqrt{N\nb}. \nonumber
\end{align}
The peak power grows as $\sqrt{\nb}$, and the full-inversion time shrinks
as $1/\sqrt{\nb}$.

\begin{figure*}[tbp]
\centering
\includegraphics[width=\textwidth]{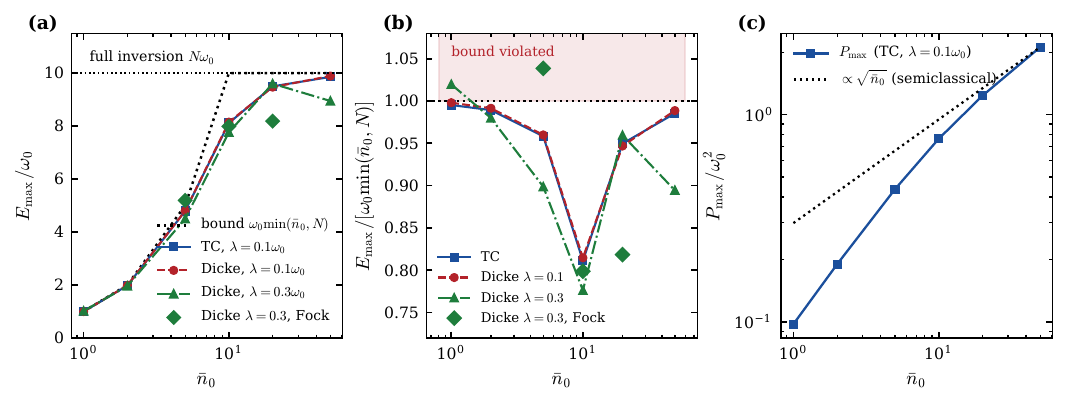}
\caption{(Color online) Summary of Sec.~\ref{sec:part1}, same parameters as
Fig.~\ref{fig:charging}.  (a) Peak stored energy
$E_{\max}/\omega_0$ versus the initial mean photon number $\nb$ (log
axis).  Dotted black: the exact TC bound $\omega_0\min(\nb,N)$ of
Eq.~\eqref{eq:bound}.  Solid blue squares: TC, $\lambda=0.1\,\omega_0$.
Dashed red circles: Dicke, $\lambda=0.1\,\omega_0$.  Dash-dotted green
up-triangles: Dicke, $\lambda=0.3\,\omega_0$.  Green diamonds: Dicke,
$\lambda=0.3\,\omega_0$, Fock reservoirs ($n=5,10,20$).
(b) The same data normalised by the bound.  The blue and red
curves stay at or below $1$; the green dash-dotted curve and the green
diamond at $n=5$ exceed it, showing that the counter-rotating terms
break the TC bound at ultrastrong coupling.
(c) Peak power $P_{\max}$ of the TC model (solid blue squares)
against the semiclassical prediction $P_{\max}\propto\sqrt{\nb}$ of
Eq.~\eqref{eq:semi} (dotted black, normalised at $\nb=50$).  The slope
approaches $1/2$ on the log-log plot, confirming the scaling; the
absolute value rises from $31\%$ of the semiclassical prediction at
$\nb=1$ to $95\%$ at $\nb=50$, the deficit being the finite-$N$
correction to Eq.~\eqref{eq:semi}.}
\label{fig:part1summary}
\end{figure*}

\subsection{Numerical results}\label{sec:part1num}

Figure~\ref{fig:charging} shows the exact dynamics obtained by
diagonalising $\hat H$ in the truncated Fock basis ($n_{\max}=110$,
$N=10$, $\lambda=0.1\,\omega_0$, $T=120$).  Panel~(a) displays
$E(t)/(N\omega_0)$ for $\nb=1,2,5,10,20,50$.  The envelope of the
curves grows monotonically with $\nb$, the peak value saturates at the
full-inversion value $N\omega_0$, and the time to reach it decreases
monotonically from
$16.1\,\omega_0^{-1}$ at $\nb=1$ to $7.4\,\omega_0^{-1}$ at $\nb=50$,
in agreement with Eq.~\eqref{eq:semi} ($7.0\,\omega_0^{-1}$ at
$\nb=50$, i.e. within $5\%$).

Panel~(b) of Fig.~\ref{fig:charging} is a direct test of
Eqs.~\eqref{eq:shortTC} and \eqref{eq:shortDicke}: we plot
$\mean{\Ne(t)}/(\lambda^2t^2)$ versus $t$ up to $t=4\,\omega_0^{-1}$
(the grid spacing is $\Delta t=0.125\,\omega_0^{-1}$).  Every curve
starts on the analytic value and bends away from it at a time of order
$1/(\lambda\sqrt{4\nb+1})$.  To make the comparison quantitative rather
than visual we extract the coefficient itself by Richardson
extrapolation: writing $\mean{\Ne(t)}/t^2=c_2+c_4t^2+c_6t^4$ and
interpolating it through the three innermost grid points,
$t=0.125,0.250,0.375\,\omega_0^{-1}$, gives $c_2/\lambda^2=\nb$ for the
TC model
to a relative accuracy better than $10^{-5}$ for all six reservoirs, and
$c_2/\lambda^2=4\nb+1$ for the Dicke model to within $0.09\%$ at
$\nb=1$ and $0.15\%$ at $\nb=50$ (Table~\ref{tab:shorttime}).  This
verifies numerically the claim that more initial photons excite more
atoms,
and it fixes the coefficient and not only the exponent.

Figure~\ref{fig:part1summary} quantifies the saturation of the bound.
Panel~(a) shows $E_{\max}/\omega_0$ against $\nb$: the TC and the weakly
coupled Dicke model ($\lambda=0.1\,\omega_0$) track the bound
$\omega_0\min(\nb,N)$ closely, reaching $9.85\,\omega_0$ and
$9.89\,\omega_0$ at $\nb=50$, i.e. $98.5\%$ and $98.9\%$ of the
$N\omega_0=10\,\omega_0$ ceiling.  Panel~(b) shows the ratio
$E_{\max}/[\omega_0\min(\nb,N)]$.  For TC and Dicke at
$\lambda=0.1\,\omega_0$ this ratio never exceeds $1$, as
Eq.~\eqref{eq:bound} requires.  For the ultrastrong Dicke coupling
$\lambda=0.3\,\omega_0$ it does: a coherent reservoir with $\nb=1$
stores $E_{\max}=1.020\,\omega_0$ ($102.0\%$ of the bound) and a
five-photon Fock reservoir stores $E_{\max}=5.193\,\omega_0$
($103.9\%$).  The excess energy is supplied by the counter-rotating
terms, which create real photons out of the virtual ones already present
in the ground state of the coupled system.  This violation of the TC
bound is a genuinely quantum feature of the Dicke battery and is the
first indication that the reservoir argument above is not the whole
story once the environment is allowed to be active.

\begin{figure*}[tbp]
\centering
\includegraphics[width=\textwidth]{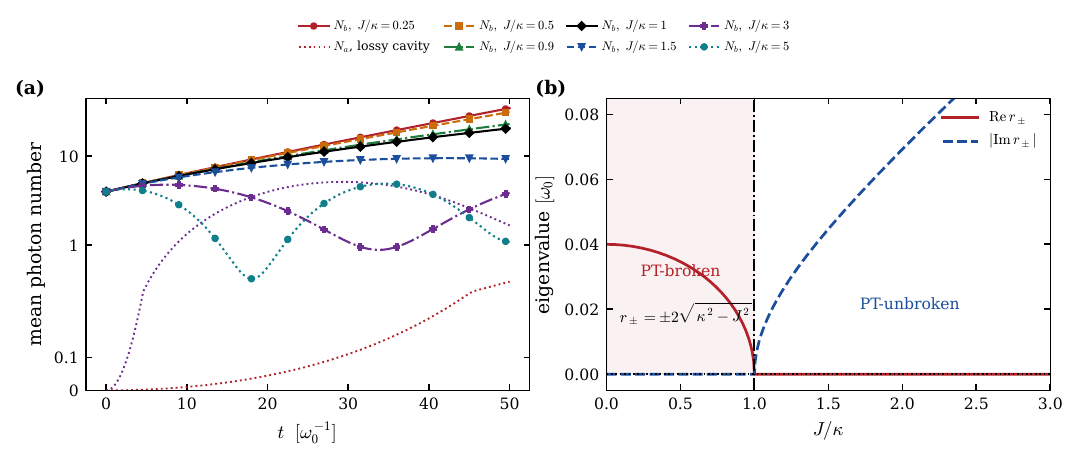}
\caption{(Color online) The $\mathcal{PT}$-symmetric pair alone,
$\gamma=0.04\,\omega_0$ ($\kappa=0.02\,\omega_0$), no battery,
$t\le50$.  (a) Mean photon numbers from the exact moment
solution.  Solid, dashed and dash-dotted lines with markers are the gain
cavity $N_b$ for $J/\kappa=0.25$ (red, circles), $0.5$ (orange,
squares), $0.9$ (green, up-triangles), $1$ (black, diamonds),
$1.5$ (purple, down-triangles), $3$ (purple, plus) and $5$ (teal,
circles); the two thin dotted lines are the lossy cavity $N_a$ for
$J/\kappa=0.25$ and $3$.  Broken-phase curves grow exponentially;
unbroken-phase curves are bounded and oscillatory, and their lossy
cavity carries a large population.
(b) Nonzero eigenvalues $r_\pm$ of the moment generator,
Eq.~\eqref{eq:charpoly}: real part (solid red), absolute imaginary part
(dashed blue).  The vertical dash-dotted line marks the exceptional
point $J=\kappa$ and the shaded region is the broken phase.}
\label{fig:ptpair}
\end{figure*}

Storing energy is not the
same as storing work.  Of the energy $E$ sitting in the battery
only the part that a cyclic unitary acting on the battery alone can
remove is recoverable, and that part is the ergotropy $W$ of
Eq.~\eqref{eq:ergotropy}, evaluated here by pairing the eigenvalues of
$\hat{\rho}_B$ in decreasing order with the ladder $\epsilon_m=\omega_0m$ in
increasing order.  We quantify the loss by the extractable fraction
taken at the peak of the charge,
$\eta^\star\equiv W(t^\star)/E_{\max}$ with
$t^\star=\arg\max_tE(t)$.  For the weakly coupled models this fraction
never falls below $91\%$: it is $99.8$, $99.5$, $97.2$, $91.3$, $95.3$
and $98.5$ per cent for the TC battery at $\nb=1,2,5,10,20,50$, and
$98.8$, $98.9$, $97.6$, $94.0$, $95.0$ and $98.9$ per cent for the Dicke
battery at $\lambda=0.1\,\omega_0$.  The dip near $\nb\simeq N=10$ is
systematic and is reproduced by Fock reservoirs ($91.2\%$ at $n=10$); it
occurs where the reservoir is just large enough to invert the whole
battery, so that the population is spread over the entire ladder instead
of piling up near its top and the state is furthest from passive.  The
ultrastrong Dicke battery is noticeably worse: with a twenty-photon Fock
reservoir only $81.4\%$ of the $8.18\,\omega_0$ stored is extractable,
the remaining $1.52\,\omega_0$ being locked by the entropy that the
counter-rotating terms generate.  Charging from a pre-existing reservoir is therefore a thermodynamically
cheap regime: the charge degrades the state only
weakly and $W_{\max}$ tracks
$E_{\max}$ to within a few per cent.  The active environment of
Sec.~\ref{sec:twobat} does not share this property, as Sec.~\ref{sec:ergotropy}
shows.


\begin{table}[tb]
\caption{Short-time coefficient $c_2$ of
$\mean{\Ne(t)}/t^2=c_2+c_4t^2+c_6t^4$, obtained by Richardson
extrapolation through the three innermost grid points,
$t=0.125,0.250,0.375\,\omega_0^{-1}$
($\Delta t=0.125\,\omega_0^{-1}$, $\lambda=0.1\,\omega_0$, $N=10$).
The analytic predictions are
$c_2=\lambda^2\nb$ (TC) and $c_2=\lambda^2(4\nb+1)$ (Dicke, coherent
reservoir).}
\label{tab:shorttime}
\begin{ruledtabular}
\begin{tabular}{ccccc}
& \multicolumn{2}{c}{TC} & \multicolumn{2}{c}{Dicke}\\
\cline{2-3}\cline{4-5}
$\nb$ & $c_2/\lambda^2$ & theory & $c_2/\lambda^2$ & theory\\
\hline
1  & 1.00000 & 1  & 4.9957  & 5   \\
2  & 2.00000 & 2  & 8.9918  & 9   \\
5  & 5.00000 & 5  & 20.980  & 21  \\
10 & 10.0000 & 10 & 40.958  & 41  \\
20 & 20.0000 & 20 & 80.907  & 81  \\
50 & 49.9999 & 50 & 200.700 & 201
\end{tabular}
\end{ruledtabular}
\end{table}

\section{The $\mathcal{PT}$-symmetric cavity pair}\label{sec:part2}

\subsection{Model}\label{sec:ptmodel}

We now consider two cavity modes, $\hat a$ (lossy) and $\hat b$ (driven), coupled
by a beam-splitter interaction.  In the rotating frame,
\begin{align}
\hat H&=J\left(\ad \hat b+\hat a\bd\right),
\label{eq:hpt}\\
\mathcal{L}\hat{\rho}&=-i[\hat H,\hat{\rho}]+\gamma\mathcal{D}[\hat a]\hat{\rho}
+\gamma\mathcal{D}[\bd]\hat{\rho}.
\label{eq:liouv}
\end{align}
with $\mathcal{D}[\hat L]\hat{\rho}=\hat L\hat{\rho} \hat L^\dagger-\tfrac12\{\hat L^\dagger \hat L,\hat{\rho}\}$ in
the standard Gorini--Kossakowski--Sudarshan--Lindblad
form~\cite{BreuerPetruccione}.  The gain channel is implemented by the
jump operator $\hat L_b=\sqrt\gamma\,\bd$, which is the standard linear
amplifier and necessarily adds the accompanying amplifier
noise~\cite{Caves1982,ScullyZubairy,Clerk2010}.  We define
$\kappa=\gamma/2$.  The Hamiltonian~\eqref{eq:hpt} is the
beam-splitter generator of passive $\mathcal{PT}$ symmetry: under
parity $\hat a\leftrightarrow\hat b$ and time reversal
$i\to-i$, $\gamma\to-\gamma$, the Liouvillian is invariant.  This
realisation differs from the semiclassical $\mathcal{PT}$ symmetry of a
driven-dissipative Kerr oscillator, where the non-Hermitian description
emerges only after eliminating quantum
fluctuations~\cite{Zhang2021}; here the loss and the gain are genuine
Lindblad channels acting on an otherwise linear system.  Engineering
balanced loss and gain is also the standard route to nonreciprocal,
directional photon transmission between two
modes~\cite{Metelmann2015}, although amplification never comes for free
once quantum noise is accounted for~\cite{Lau2018}.

\subsection{Exact moment dynamics}\label{sec:moments}

Because the Hamiltonian is quadratic and both jump operators are linear
in the mode operators, the adjoint Liouvillian
$\mathcal{L}^\dagger[\hat O]=i[\hat H,\hat O]+\gamma\mathcal{D}^\dagger[\hat a][\hat O]
+\gamma\mathcal{D}^\dagger[\bd][\hat O]$ maps the space of operator
polynomials of degree $\le2$ into itself.  The first and second moments
therefore obey a closed linear system; no truncation or
Gaussian assumption is involved.  Appendix~\ref{app:moments} derives the
reduced three-dimensional subsystem for
\begin{equation}
N_a=\mean{\ad a},\qquad N_b=\mean{\bd b},\qquad
S=i\mean{\ad b-a\bd},
\end{equation}
namely
\begin{subequations}\label{eq:moment3}
\begin{align}
\dot N_a&=-J S-\gamma N_a,\label{eq:m3a}\\
\dot N_b&=+J S+\gamma\left(N_b+1\right),\label{eq:m3b}\\
\dot S&=-2J\left(N_b-N_a\right).\label{eq:m3c}
\end{align}
\end{subequations}
The inhomogeneous term $\gamma$ in Eq.~\eqref{eq:m3b} is the spontaneous
emission of the amplifier: a linear gain channel populates even the
vacuum.  The characteristic polynomial of
Eq.~\eqref{eq:moment3} is
\begin{equation}
-r\left(r^2-\gamma^2+4J^2\right)=0
\label{eq:charpoly}
\end{equation}
so the eigenvalues are $0$ and
$r_\pm=\pm2\sqrt{\kappa^2-J^2}$.  These are eigenvalues of the moment
generator, the $3\times3$ matrix behind Eq.~\eqref{eq:moment3} that advances $N_a$,
$N_b$ and $S$, so $r_\pm$ are rates for the photon numbers rather than for the
Hamiltonian.  The factor of two has a simple origin.  The gain-mode amplitude
$\langle\hat b\rangle$ follows an effective non-Hermitian Hamiltonian with eigenvalues
$\pm\sqrt{J^2-\kappa^2}$, and for $J<\kappa$ it grows as $\exp(\sqrt{\kappa^2-J^2}\,t)$.
Since the photon number is the square of the amplitude, $N_b\sim|\langle\hat b\rangle|^2$,
it grows at twice that rate, which is why $r_\pm=\pm2\sqrt{\kappa^2-J^2}$.

The transition occurs at $J=\kappa$, where the two roots coalesce.  In
the broken phase, $J<\kappa$, $r_\pm$ are real and
$N_b$ grows without bound at the rate
$2\sqrt{\kappa^2-J^2}$; as $J\to0$ this reduces to the bare
amplifier rate $\gamma$.  In the unbroken phase, $J>\kappa$,
$r_\pm$ are purely
imaginary and the photon numbers oscillate.  Because of the
zero eigenvalue, however, the solution is not strictly periodic: the
secular mode grows linearly with the slope
\begin{equation}
v=\frac{\gamma J^2}{2\left(J^2-\kappa^2\right)}
\;\xrightarrow[J\to\infty]{}\;\frac{\gamma}{2},
\label{eq:secular}
\end{equation}
obtained from the left and right null vectors of the generator
(Appendix~\ref{app:moments}).  The unbounded but only linear
growth of the unbroken phase is therefore entirely attributable to the
spontaneous term of the amplifier; the coherent part of the dynamics is
bounded.

\subsection{Verification}\label{sec:ptverify}

We verify Eq.~\eqref{eq:moment3} in two independent ways.  First, we
construct
the full $15$-dimensional moment vector
$\{1,\;\hat a,\hat b,\ad,\bd,\;\hat a\ad,\bd \hat b,\ad \hat b,\hat a\bd,\;\hat a^2,\hat b^2,\ad{}^2,\bd{}^2,\hat a\hat b,\ad\bd\}$
and fit the adjoint Liouvillian on it by least squares on a truncated
Fock space with $n_{\max}=90$ (closure residual $<10^{-10}$).  The
resulting trajectories agree with Eq.~\eqref{eq:moment3} to
$\max_t|\Delta N_{a,b}|<2.5\times10^{-12}$ simultaneously for coherent
($\nb=4$), vacuum and Fock ($n=5$) initial states at
$J/\kappa=0.25,0.5,0.9,1,1.5,3,5$.  Second, with the same
generator, the moment solution agrees with a direct vectorised
propagation of the master equation~\eqref{eq:liouv} to
$1.8\times10^{-4}$ over $t\le50$, and with $K=400$ Monte-Carlo
wave-function trajectories~\cite{Dalibard1992,Dum1992,Molmer1993,Plenio1998}
to $0.037$ in $N_a$ and $0.093$ in $N_b$, consistent with the expected
statistical error $\sim K^{-1/2}$.

Figure~\ref{fig:ptpair} shows the result.  In panel~(a) the broken
cases ($J/\kappa=0.25,0.5,0.9$) grow monotonically and reach
$N_b=34.8$, $31.5$ and $23.1$ at $t=50$ from an initial $\nb=4$, while
the unbroken cases stay bounded and oscillatory: $N_b$ peaks at $4.8$
for $J/\kappa=3$ and at $4.9$ for $J/\kappa=5$, returning to $1.1$ by
$t=50$ in the latter case.  The lossy cavity carries almost nothing in
the broken phase ($\max_t N_a=0.39$ at $J/\kappa=0.25$) but a
comparable amount in the unbroken phase ($\max_t N_a=5.2$ at
$J/\kappa=3$).  Panel~(b) plots the analytic eigenvalues
$r_\pm$; the exceptional point is at $J/\kappa=1$, where the real
and imaginary branches meet with the characteristic square-root
coupling.

\begin{figure*}[tbp]
\centering
\includegraphics[width=\textwidth]{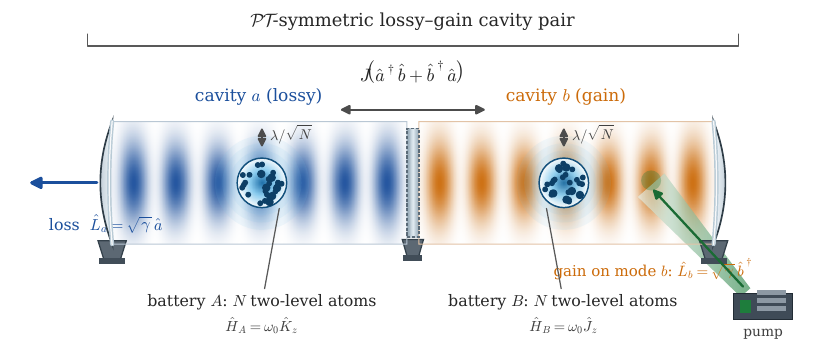}
\caption{(Color online) Model.  Two identical cavities are coupled by the
beam-splitter interaction of strength $J$; the left cavity $\hat a$ is lossy
with rate $\gamma$, the right cavity $\hat b$ is amplifying with the same rate,
and together they form a $\mathcal{PT}$-symmetric pair whose exceptional
point lies at $J=\kappa=\gamma/2$.  An identical battery of $N$
two-level atoms sits in each cavity and couples to its mode with strength
$\lambda/\sqrt N$; the gain-side battery $B$ has
$\hat H_B=\omega_0\jz$, the lossy-side battery $A$ has
$\hat H_A=\omega_0\kz$, and both start in the collective ground state
$\ket{j,-j}$.  The standing-wave
intensity of each mode is sketched inside the cavities; the brighter shade
of cavity $\hat b$ indicates the photon pile-up in the broken phase, and the
pump and leakage channels are drawn as arrows.  In the broken phase
($J<\kappa$) the gain fills cavity $\hat b$ faster than the pair can empty it,
so the reservoir seen by battery $B$ keeps growing while battery $A$ sees
almost nothing; in the unbroken phase
($J>\kappa$) photons are shuttled into $\hat a$, where battery $A$ can
intercept them before they are dissipated.}
\label{fig:model}
\end{figure*}

We now attach the battery, and to keep the $\mathcal{PT}$ symmetry of the
pair intact one places an identical battery in each cavity, as sketched in
Fig.~\ref{fig:model}.

\section{Two batteries and strict PT symmetry}\label{sec:twobat}

\subsection{Model and observables}\label{sec:symmodel}

With an identical battery in each cavity the Hamiltonian becomes
\begin{align}
\hat H&=J\left(\ad \hat b+\hat a\bd\right)
+\frac{\lambda}{\sqrt N}\left(\hat b\,\jp+\bd\,\jm\right)\nonumber\\
&\quad+\frac{\lambda}{\sqrt N}\left(\hat a\,\kp+\ad\,\km\right),
\label{eq:symH}
\end{align}
again in the rotating frame.  The free energies
$\omega(\ad\hat a+\bd\hat b)+\omega_0(\jz+\kz)$ are removed by the frame
change of Sec.~\ref{sec:models}, and for this coupling the removal is
exact rather than approximate: the interaction conserves the total
excitation number, $[\hat n_a+\hat n_b+\jz+\kz,\hat H]=0$, so the free
terms generate a phase and nothing else, and the laboratory frame
returns the same $\mean{\hat n_a}$, $\mean{\hat n_b}$ and stored
energies.

with the same Lindblad operators $\hat L_a=\sqrt\gamma\,\hat a$ and
$\hat L_b=\sqrt\gamma\,\bd$ as before; here $\hat J$ and $\hat K$ are the
collective spins of the batteries in the gain and the lossy cavity.
The Liouvillian is now strictly $\mathcal{PT}$-symmetric.  The parity
operation exchanges the two cavities and the two spins, $a\leftrightarrow b$
and $J\leftrightarrow K$, and time reversal sends $\hat a\to\ad$,
$\hat b\to\bd$, $\jp\leftrightarrow\jm$ and $\kp\leftrightarrow\km$.  The
beam-splitter term and both atom--field terms are invariant under each of
these maps, while the two jump operators are simply interchanged, so the
full generator is $\mathcal{PT}$-invariant.

The initial state, however, is not symmetric: it holds $\nb=4$ photons in
the gain cavity and none in the lossy cavity.  The symmetry of the
generator therefore does not force the two batteries to behave alike, and
the directionality of the broken phase survives.

Loading the cavities leaves the exceptional point at $J=\kappa$, and
Sec.~\ref{sec:ep} shows why: in the low-excitation limit the coupling
$\lambda$ drops out of the branch-point condition altogether.  The
spectrum is not what it was, however.  At $J=\kappa$ the critical
eigenvalues are $\pm i\lambda$ rather than zero, and throughout the
broken phase the growth rate is half that of the empty pair.  Both
statements belong to linear response; a battery charged to
$\mean{\Ne}\simeq5$ out of $N=10$ is past that regime, so the
exceptional point is to be read as the boundary of the linear
instability and not as a property of the charging curves
themselves.

We use three observables, defined for the gain-side battery and repeated
for the lossy-side one.  The stored energy
$E_B(t)=\omega_0\mean{\Ne^B(t)}$ with
$\Ne^B=\jz+N/2$ the number of excitations in battery $B$, and
$E_A(t)=\omega_0\mean{\Ne^A(t)}$ with $\Ne^A=\kz+N/2$; the charging
power, evaluated as the
equal-time operator
\begin{align}
P(t)&=-2\,\frac{\lambda}{\sqrt N}\,\omega_0\,\mathrm{Im}\mean{\bd\,\jm(t)}
\nonumber\\
&=-\frac{2\lambda\omega_0}{\sqrt N}\,\mean{\hat O_P},\qquad
\hat O_P\equiv\frac{\bd\,\jm-\hat b\,\jp}{2i},
\label{eq:power}
\end{align}
so that no numerical differentiation of a stochastic trajectory is
required; and the ergotropy
\begin{equation}
W(t)=\Tr[\hat{\rho}_B(t)\hat H_B]-E_{\rm passive},
\qquad \hat H_B=\omega_0\jz,
\label{eq:ergotropy}
\end{equation}
where $E_{\rm passive}=\sum_i p_i^{\downarrow}\epsilon_i^{\uparrow}$ is
obtained by pairing the eigenvalues of $\hat{\rho}_B$ in decreasing order with
the eigenvalues $\epsilon_m=\omega_0 m$ of $\hat H_B$ in increasing
order~\cite{Pusz1978,Allahverdyan2004}.  The inequality $W\le E$ follows
by construction, and imposing it is the consistency check we run on
every trajectory set.  Battery $A$ is measured with the same expressions
after the replacement $\hat b\to\hat a$ and $\hat J\to\hat K$, which gives
the columns $E_A$, $P_A$ and $W_A$ of Table~\ref{tab:twobat}.

\subsection{Two-sided charging}\label{sec:twosided}

\begin{figure*}[tbp]
\centering
\includegraphics[width=\textwidth]{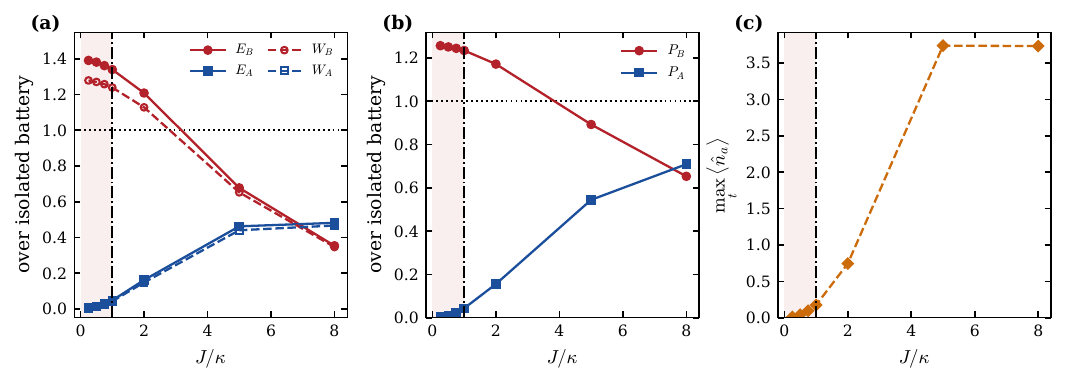}
\caption{(Color online) Two-sided charging across the exceptional point.
(a) Peak stored energy (solid lines, filled markers) and ergotropy
(dashed lines, open markers) of the gain-side battery $B$ (circles) and
the lossy-side battery $A$ (squares), each normalised to the isolated
battery.  (b) Peak charging power, same normalisation.
(c) Maximum population ever carried by the lossy cavity,
$\max_t\mean{\hat n_a}$.  In each panel the vertical dash-dotted line
marks the exceptional point and the shaded band is the broken phase.
Parameters $N=10$, $\omega_0=1$, $\lambda=0.1$, $\gamma=0.04$.}
\label{fig:twobat}
\end{figure*}

Figure~\ref{fig:twobat}
and Table~\ref{tab:twobat} collect the results, computed with
$N=10$, $\omega_0=1$, $\lambda=0.1$, $\gamma=0.04$ and the Monte-Carlo method
of Sec.~\ref{sec:method}.

\begin{table*}[!tbp]
\caption{Two-sided charging.  Peak stored energy $E$, charging power $P$
and ergotropy $W$ of the gain-side battery $B$ and the lossy-side battery
$A$ in the strictly $\mathcal{PT}$-symmetric setup of
Eq.~\eqref{eq:symH}.  Parameters $N=10$, $\omega_0=1$, $\lambda=0.1$,
$\gamma=0.04$; energies
and ergotropies are in units of $\omega_0$, powers in units of
$\omega_0^2$.  The statistical uncertainty is $\sim2$--$3\%$ (see
Sec.~\ref{sec:method}).}
\label{tab:twobat}
\begin{ruledtabular}
\begin{tabular}{ccccccc}
$J/\kappa$ & $E_B$ & $E_A$ & $P_B$ & $P_A$ & $W_B$ & $W_A$\\
\colrule
0.25 & $5.41$ & $0.01$ & $0.45$ & $0.00$ & $4.89$ & $0.01$\\
0.5 & $5.37$ & $0.05$ & $0.45$ & $0.00$ & $4.86$ & $0.04$\\
0.75 & $5.29$ & $0.11$ & $0.45$ & $0.01$ & $4.81$ & $0.09$\\
1.0 & $5.21$ & $0.18$ & $0.44$ & $0.02$ & $4.74$ & $0.16$\\
2.0 & $4.70$ & $0.63$ & $0.42$ & $0.06$ & $4.31$ & $0.56$\\
5.0 & $2.63$ & $1.80$ & $0.32$ & $0.20$ & $2.49$ & $1.68$\\
8.0 & $1.37$ & $1.87$ & $0.23$ & $0.25$ & $1.32$ & $1.78$\\
\end{tabular}
\end{ruledtabular}
\end{table*}

In the broken phase the gain-side battery $B$ charges while the
lossy-side battery $A$ stays empty: at $J/\kappa=0.25$,
$E_B=5.41\,\omega_0$ against $E_A=0.01\,\omega_0$.  The gain
piles photons into $\hat b$, the loss drains $\hat a$, and the energy is
delivered to $B$ alone.  Across the exceptional point this picture
changes direction.  Deep in the unbroken phase at $J/\kappa=8$ the two
batteries share what energy survives the loss, $E_B=1.37\,\omega_0$ and
$E_A=1.87\,\omega_0$, with $A$ now ahead of $B$: the strong coupling
shuttles photons from $\hat b$ to $\hat a$, and battery $A$ sits at the
downstream end of this current, just before the loss, where it intercepts
the photons on their way out.  The power and the ergotropy follow the
same pattern.

Figure~\ref{fig:twobat} is the main result.  It shows the peak stored
energy, peak power and ergotropy of both batteries, as $J/\kappa$ is
scanned across the exceptional
point.  Battery $B$ gains $39\%$ in energy, $26\%$ in power and $28\%$ in
ergotropy at $J/\kappa=0.25$, its best case, and loses $65\%$, $35\%$ and
$65\%$ at $J/\kappa=8$.  Battery $A$ never beats the isolated battery,
but it loses far less in the unbroken phase, $52\%$, $29\%$ and $53\%$ at
$J/\kappa=8$, and its share grows monotonically with $J$; the crossover
at which $E_A$ overtakes $E_B$ lies between $J/\kappa=5$ and $8$.  The
sum $E_A+E_B$ declines much more slowly than $E_B$ alone, from
$5.42\,\omega_0$ at $J/\kappa=0.25$ to $3.24\,\omega_0$ at $8$: what the
current removes from $\hat b$ it partly delivers to $\hat a$.

Restoring the strict $\mathcal{PT}$ symmetry of the generator therefore
does not remove the advantage of the broken phase.  The directionality
that makes a $\mathcal{PT}$-broken pair a good charger comes from the
choice of initial state, the photons placed in the gain cavity, rather
than from any asymmetry of the Liouvillian itself.

The rest of this section compares the two batteries in detail: their time
traces, the mechanism that decides which of them receives the energy, and
how much of either charge can be recovered as work.

\subsection{Dynamics of the two batteries}\label{sec:fullresults}

\begin{figure*}[tbp]
\centering
\includegraphics[width=\textwidth]{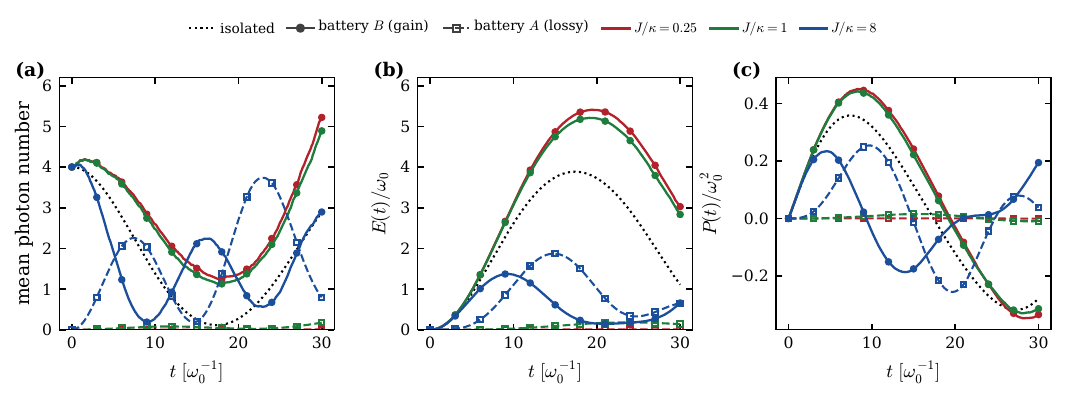}
\caption{(Color online) Time traces of the two batteries,
$N=10$, $\lambda=0.1\,\omega_0$, $\gamma=0.04\,\omega_0$
($\kappa=0.02\,\omega_0$), $\nb=4$, $T=30$, $K=300$ trajectories.
Battery $B$ in the gain cavity is drawn solid with filled markers and
battery $A$ in the lossy cavity dashed with open markers; the coupling is
coded by colour, red for $J/\kappa=0.25$ (broken), green for
$J/\kappa=1$ and blue for $J/\kappa=8$ (unbroken), and the black dotted
line is the isolated battery.  (a) Mean photon number of the gain
cavity $\hat b$ and of the lossy cavity $\hat a$.
(b) Stored energy $E(t)/\omega_0$.
(c) Charging power $P(t)/\omega_0^2$, evaluated with the
equal-time operator~\eqref{eq:power}; negative values correspond to
discharging.}
\label{fig:fulltraces}
\end{figure*}

Figure~\ref{fig:fulltraces} follows both batteries in time for three
couplings spanning the two phases; the isolated battery is shown for
reference.

Panel~(a) displays the mean photon numbers.  In the broken phase
($J/\kappa=0.25$) the gain cavity fills monotonically while the lossy
cavity stays nearly empty, and battery $B$ charges alone; in
the unbroken phase the two cavities exchange population rapidly, and
for $J/\kappa=8$ the lossy cavity carries up to $3.7$ photons, enough
for battery $A$ to overtake $B$.  Panels~(b) and~(c)
show the corresponding stored energy and power; the two batteries peak
at different times, $A$ later than $B$, which is the first sign that the
energy reaching $A$ has already passed through $\hat b$.

\subsection{Mechanism}\label{sec:mechanism}

Two effects cooperate, and they can be separated by looking at where
the photons are.

In the broken phase the coherent exchange cannot keep up with the
amplification, so the
photons pile up in $\hat b$; battery $B$ is the only sink available there,
and they are absorbed by it.  Battery $A$ sits on the other side of a
coupling it cannot beat, and stays empty.  The total excitation number
$\mean{\hat n_a+\hat n_b+\Ne^A+\Ne^B}$, exactly conserved for the isolated
battery and something the gain can only increase, reaches $8.3$ at
$J/\kappa=0.25$ versus $4.0$ for the isolated battery, i.e. the
environment has injected $4.3$ excitations.  Since the TC bound of
Eq.~\eqref{eq:bound} caps an isolated battery at $\nb=4$, this
injection is what allows $E_B$ to exceed the isolated
value.  This is the physical picture sketched in the
Introduction: the $\mathcal{PT}$-broken environment does for battery $B$
what a larger initial reservoir did in Sec.~\ref{sec:part1}.

In the unbroken phase the
two cavities hybridise into normal modes at $\pm J$ and exchange
population coherently.  Photons that reach $\hat a$ meet battery $A$
first and are dissipated at rate $\gamma$ once they pass it.
Panel~(c) of
Fig.~\ref{fig:twobat} makes this quantitative: $\max_t\mean{\hat n_a}$ is
$0.011$ at $J/\kappa=0.25$, $0.042$ at $0.5$, $0.096$ at $0.75$ and
$0.175$ at $1$, then grows steeply to $0.74$ at $J/\kappa=2$ and
$3.74$ at $5$, where it saturates.  Battery $A$ intercepts this flux
before it reaches the drain, its stored energy
rising monotonically with $J$ while $B$ loses
$4.0\,\omega_0$ over the same range.  Concomitantly the
total excitation number at $t=T$ falls from $8.3$ to $5.0$.  The
hybridisation also detunes the normal modes from the atomic transition
by $\pm J$, which suppresses the battery coupling in second order;
both effects act in the same direction, and together they produce the
steep monotone decrease seen in Fig.~\ref{fig:twobat}(a).

Taken together, the two effects confirm the picture sketched in the
Introduction, with one
addition.  The deep unbroken phase actively hurts: a
$\mathcal{PT}$-symmetric pair whose coupling dominates the loss and gain
rate acts, as far as the battery is concerned, as an efficient leak.
The best operating point therefore sits on the broken side, a little
below the exceptional point.

\subsection{Ergotropy and the locked energy}\label{sec:ergotropy}

The stored energy is an upper bound on what either battery can deliver.
A cyclic unitary acting on the battery
alone recovers the ergotropy $W$ of Eq.~\eqref{eq:ergotropy}, and the
difference
\begin{equation}
L(t)\equiv E(t)-W(t)\ge0
\label{eq:locked}
\end{equation}
is locked energy: it sits in the battery, but the eigenvalues of
its reduced state are not ordered oppositely to the spectrum of its
$\hat H$, so no
unitary on the battery alone can reach it~\cite{Pusz1978,Allahverdyan2004}.
We evaluate $W$ for both batteries over every run of the $J$ scan, as a
function of time
as well as at its peak values, and we separate the locked energy
$L$ into a quantum and a classical part.  All numbers are
collected in Table~\ref{tab:ergotropy}.

\begin{table*}[!tbp]
\caption{Ergotropy of the two batteries across the exceptional point.
$E_{\max}$ and $W_{\max}$ are the peak stored energy and ergotropy;
$\eta=W_{\max}/E_{\max}$; $\eta^\star=W(t^\star)/E_{\max}$ at the charge
peak $t^\star=\arg\max_t E$, so that $1-\eta^\star$ is the locked
fraction then; $L_{\max}=\max_t[E(t)-W(t)]$.  The statistical
uncertainty is
$\sim2$--$3\%$ (see Sec.~\ref{sec:method}).}
\label{tab:ergotropy}
\begin{ruledtabular}
\begin{tabular}{lcccccc}
battery & $J/\kappa$ & $E_{\max}/\omega_0$ & $W_{\max}/\omega_0$
 & $\eta$ & $1-\eta^\star$ & $L_{\max}/\omega_0$\\
\colrule
isolated & --- & $3.886$ & $3.820$ & $0.983$ & $1.7\%$ & $0.112$\\
\colrule
$B$ & $0.25$ & $5.407$ & $4.886$ & $0.904$ & $11.2\%$ & $1.608$\\
$B$ & $0.5$ & $5.370$ & $4.855$ & $0.904$ & $10.4\%$ & $1.533$\\
$B$ & $0.75$ & $5.294$ & $4.806$ & $0.908$ & $10.2\%$ & $1.493$\\
$B$ & $1$ & $5.209$ & $4.735$ & $0.909$ & $10.1\%$ & $1.440$\\
$B$ & $2$ & $4.698$ & $4.311$ & $0.918$ & $8.9\%$ & $1.215$\\
$B$ & $5$ & $2.632$ & $2.491$ & $0.946$ & $5.6\%$ & $0.386$\\
$B$ & $8$ & $1.373$ & $1.322$ & $0.963$ & $3.8\%$ & $0.400$\\
\colrule
$A$ & $0.25$ & $0.012$ & $0.011$ & $0.876$ & $14.5\%$ & $0.005$\\
$A$ & $0.5$ & $0.048$ & $0.042$ & $0.879$ & $13.5\%$ & $0.018$\\
$A$ & $0.75$ & $0.106$ & $0.093$ & $0.882$ & $13.4\%$ & $0.038$\\
$A$ & $1$ & $0.183$ & $0.162$ & $0.883$ & $13.0\%$ & $0.066$\\
$A$ & $2$ & $0.627$ & $0.564$ & $0.899$ & $10.8\%$ & $0.206$\\
$A$ & $5$ & $1.796$ & $1.681$ & $0.936$ & $6.6\%$ & $0.375$\\
$A$ & $8$ & $1.872$ & $1.782$ & $0.952$ & $4.8\%$ & $0.398$\\
\end{tabular}
\end{ruledtabular}
\end{table*}

\begin{table*}[!tbp]
\caption{Decomposition of the locked energy at the charge maximum
$t^\star$, from the $K=300$ runs with per-trajectory reduced states.
Fractions are relative to that battery's stored energy at that instant;
the last two columns are the maxima the two pieces reach at any time.
Battery $A$ charges too little in the broken phase for its percentages
to be meaningful, so only cases with $E_A>0.5\,\omega_0$ are listed.}
\label{tab:split}
\begin{ruledtabular}
\begin{tabular}{lccccc}
battery & $J/\kappa$ & $L_{\rm ent}/E$ & $L_{\rm cl}/E$ & $L/E$
 & $\max_t L_{\rm ent}$, $\max_t L_{\rm cl}$\\
\colrule
$B$ & $0.25$ & $3.5\%$ & $7.6\%$ & $11.2\%$ & $0.725$, $0.883$\\
$B$ & $0.75$ & $3.1\%$ & $7.1\%$ & $10.2\%$ & $0.674$, $0.820$\\
$B$ & $2$ & $2.5\%$ & $6.4\%$ & $8.9\%$ & $0.556$, $0.660$\\
$B$ & $5$ & $1.1\%$ & $4.5\%$ & $5.6\%$ & $0.185$, $0.201$\\
\colrule
$A$ & $2$ & $3.7\%$ & $7.1\%$ & $10.8\%$ & $0.089$, $0.117$\\
$A$ & $5$ & $1.5\%$ & $5.0\%$ & $6.6\%$ & $0.165$, $0.210$\\
\end{tabular}
\end{ruledtabular}
\end{table*}

\begin{figure*}[tbp]
\centering
\includegraphics[width=\textwidth]{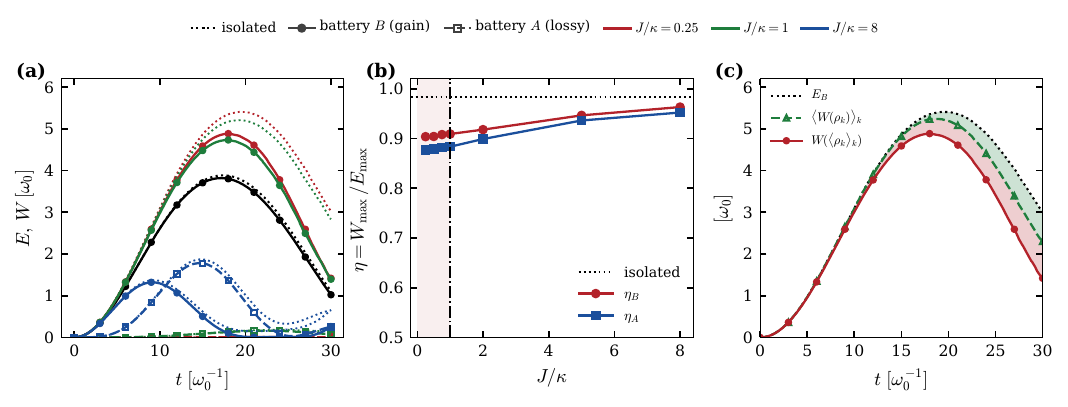}
\caption{(Color online) Ergotropy of the two batteries,
same parameters and the same colour code as Fig.~\ref{fig:fulltraces}.
(a) Ergotropy $W(t)$ (curves with markers; battery $B$ solid with filled
circles, battery $A$ dashed with open squares) together with the stored
energy $E(t)$ (dotted curves) for the isolated battery and for
$J/\kappa=0.25$, $1$ and $8$; the vertical distance between the dotted
curve and the marker curve of a given colour and battery is the locked
energy of Eq.~\eqref{eq:locked}.
(b) Extractable fraction $\eta=W_{\max}/E_{\max}$ of battery $B$ (red
circles) and battery $A$ (blue squares) versus $J/\kappa$.  The dotted
horizontal line is the isolated-battery value $\eta=0.983$, the vertical
dash-dotted line is the exceptional point and the shaded band is the
broken phase.
(c) Decomposition of the locked energy of battery $B$ at $J/\kappa=0.25$:
stored energy $E_B$ (dotted black), trajectory-averaged ergotropy
$\langle W(\hat{\rho}_k)\rangle_k$ (dashed green up-triangles) and ergotropy of
the averaged state $W(\langle\hat{\rho}_k\rangle_k)$ (solid red circles).  The
green shaded area is the entanglement-locked part $L_{\rm ent}$ and the
red shaded area the classically locked part $L_{\rm cl}$ of
Eq.~\eqref{eq:split}.}
\label{fig:ergotropy}
\end{figure*}

The environment trades quantity against quality for both batteries.
Figure~\ref{fig:ergotropy}(a) shows
that $W(t)$ follows $E(t)$ closely in every phase: the two curves have
the same shape and peak at nearly the
same time, so the environment changes mainly how much energy gets
stored.  It changes the quality of that energy too, and does so
monotonically.  Panel~(b) and Table~\ref{tab:ergotropy} show
that the extractable fraction $\eta$ of battery $B$ is minimal on the
broken side,
$0.904$ at $J/\kappa=0.25$, and rises monotonically to $0.963$ at
$J/\kappa=8$; the isolated battery sits at $0.983$, and battery $A$
runs about half a per cent below $B$ throughout, from $0.876$ in the
broken phase to $0.952$ at $J/\kappa=8$.  The locked fraction
$1-\eta^\star$, taken at that same instant, tells the story more sharply:
it is $1.7\%$ for the isolated battery, $10$--$11\%$ for $B$ throughout
the broken phase and just above the exceptional point, and falls to
$3.8\%$ at $J/\kappa=8$.  A gain channel is a source of entropy as well
as of energy, and the battery pays for the extra photons with a more
mixed state.

The trade-off is nevertheless favourable in the broken phase.  Going
from the isolated battery to $J/\kappa=0.25$ adds
$\Delta E_{\max}=1.52\,\omega_0$ of stored energy and
$\Delta W_{\max}=1.07\,\omega_0$ of extractable work to battery $B$, i.e.
about seventy per cent of the extra energy survives as extra work.  On
the unbroken side the opposite happens and there is nothing to compensate
for it: at $J/\kappa=8$ battery $B$ loses
$\Delta E_{\max}=-2.51\,\omega_0$ of energy and
$\Delta W_{\max}=-2.50\,\omega_0$ of work, so nearly the whole deficit is
in the work.  Battery $A$ loses less, $-2.01$ and $-2.04\,\omega_0$, and
at $J/\kappa=8$ it holds $35\%$ more ergotropy than $B$.  Across the
exceptional point $W_B$ changes by the factor $4.89/1.32=3.7$.

The locked energy comes from the convexity of the ergotropy.  Writing
$E_{\rm passive}(\hat{\rho})=\min_{\hat U}\Tr(\hat U\hat{\rho} \hat U^\dagger \hat H_B)$,
a minimum of linear functionals of $\hat{\rho}$, we see that $E_{\rm passive}$
is concave and therefore that $W(\hat{\rho})=\Tr(\hat{\rho} \hat H_B)-E_{\rm passive}(\hat{\rho})$ is
convex.  This matters because the Monte-Carlo wave-function method
delivers the battery state as an average over trajectories,
$\hat{\rho}_B(t)=K^{-1}\sum_k\hat{\rho}_k(t)$, and convexity gives the exact
inequality
\begin{equation}
W_{\rm avg}\equiv W\!\left(\langle\hat{\rho}_k\rangle_k\right)
\le \big\langle W(\hat{\rho}_k)\big\rangle_k\equiv W_{\rm traj},
\label{eq:convex}
\end{equation}
where $\langle\cdot\rangle_k$ denotes the average over the $K$
trajectories.  The two sides have different operational meanings:
$W_{\rm avg}$ is what an experimenter who ignores the jump record can
extract, whereas
$W_{\rm traj}$ is what could be extracted if the record were read
out and a conditional unitary applied to each trajectory separately.
Because $E=\Tr(\hat{\rho}_BH_B)$ is linear the two differ only through the
passive-state term, and Eq.~\eqref{eq:convex} lets us split the locked
energy~\eqref{eq:locked} into two non-negative pieces,
\begin{equation}
L=E-W_{\rm avg}
=\big[E-W_{\rm traj}\big]
+\big[W_{\rm traj}-W_{\rm avg}\big].
\label{eq:split}
\end{equation}
\begin{equation}
L_{\rm ent}=E-W_{\rm traj},\qquad
L_{\rm cl}=W_{\rm traj}-W_{\rm avg}.
\label{eq:splitdef}
\end{equation}
Here $L_{\rm ent}=\langle\Tr(\hat{\rho}_k\hat H_B)-W(\hat{\rho}_k)\rangle_k$ is the
average of the locked energy of the individual trajectories; since each
$\hat{\rho}_k$ is the reduced state of a pure trajectory it is mixed
only because the battery is entangled with the cavities, so $L_{\rm ent}$
is genuinely entanglement-locked.  The remainder $L_{\rm cl}$ is locked
by the classical mixture over trajectories, that is, by the unread jump
record, which
superposes states that would have peaked at different times.

Figure~\ref{fig:ergotropy}(c) and Table~\ref{tab:split} show the result
for the broken-phase optimum.  At $J/\kappa=0.25$, taken at the peak of
the charge, the stored energy $E_B=5.41\,\omega_0$ splits into
$W_{\rm traj}=5.22\,\omega_0$ and $W_{\rm avg}=4.81\,\omega_0$, so that
$L_{\rm ent}=0.19\,\omega_0$ ($3.5\%$ of $E$) and
$L_{\rm cl}=0.41\,\omega_0$ ($7.6\%$ of $E$).  The
inequality~\eqref{eq:convex} is satisfied at every time and the split is
stable: over the whole trace
$L_{\rm ent}$ and $L_{\rm cl}$ never exceed $0.73\,\omega_0$ and
$0.88\,\omega_0$.  The same analysis at $J/\kappa=5$ gives $1.1\%$ and
$4.5\%$ for $B$ and $1.5\%$ and $5.0\%$ for $A$.  The classical part
dominates everywhere: about two thirds of the locked energy in the broken
phase
comes from the dephasing produced by averaging over
trajectories whose jumps occurred at different times, and not from
entanglement at all.  And the
entanglement-locked part stays small throughout, never more than about
three per cent of the stored energy of $B$, so the battery--field
correlations
built up during charging are only weakly detrimental to work extraction.
$W_{\rm traj}$ is not an achievable
gain: reading out the jump record and applying a trajectory-selective
unitary is a measurement-based protocol whose cost must be subtracted;
we use it only as an upper bound separating the two locking mechanisms.

\section{Dicke batteries in the same pair}\label{sec:dicke}

Everything in Sec.~\ref{sec:twobat} was computed with the
Tavis--Cummings coupling of Eq.~\eqref{eq:htc}, which keeps the
rotating terms only.  The Dicke coupling of Eq.~\eqref{eq:hdic} keeps
the counter-rotating ones as well, and in Sec.~\ref{sec:part1} it was
the Dicke model that behaved differently: it has no conserved
excitation number, and at ultrastrong coupling it violates the
bound~\eqref{eq:bound}.  Nothing guarantees that the two-battery
results survive that change, and this section repeats the comparison
with the counter-rotating terms restored.

\subsection{Model and the frame}\label{sec:dickemodel}

Replacing Eq.~\eqref{eq:htc} by Eq.~\eqref{eq:hdic} in each cavity adds
$\frac{\lambda}{\sqrt N}(\bd\jp+\hat b\jm)$ to the battery in the gain
cavity and $\frac{\lambda}{\sqrt N}(\ad\kp+\hat a\km)$ to the one in
the lossy cavity.  Both are off resonant by $2\omega_0$, and that
detuning decides everything: it is what makes them a small correction
here and a large one in the ultrastrong regime of
Sec.~\ref{sec:part1}.  In the frame rotating at $\omega_0$, the frame
of Eq.~\eqref{eq:symH}, they are time dependent,
\begin{align}
\hat H(t)&=J\left(\ad\hat b+\hat a\bd\right)\nonumber\\
&\quad+\frac{\lambda}{\sqrt N}\Bigl(\hat b\jp+\bd\jm\nonumber\\
&\qquad+e^{-2i\omega_0 t}\,\hat b\jm
+e^{2i\omega_0 t}\,\bd\jp\Bigr)\nonumber\\
&\quad+\frac{\lambda}{\sqrt N}\Bigl(\hat a\kp+\ad\km\nonumber\\
&\qquad+e^{-2i\omega_0 t}\,\hat a\km
+e^{2i\omega_0 t}\,\ad\kp\Bigr),
\label{eq:dickeH}
\end{align}
with the Lindblad operators $\hat L_a=\sqrt\gamma\,\hat a$ and
$\hat L_b=\sqrt\gamma\,\bd$ unchanged; in the laboratory frame the same
model reads $\omega(\ad\hat a+\bd\hat b)+\omega_0(\jz+\kz)$ together
with the beam-splitter term and
$\frac{\lambda}{\sqrt N}(\hat b+\bd)(\jp+\jm)
+\frac{\lambda}{\sqrt N}(\hat a+\ad)(\kp+\km)$.

The generator remains strictly $\mathcal{PT}$-symmetric.  Parity
exchanges the cavities and the spins; time reversal sends
$\hat a\to\ad$, $\hat b\to\bd$, $\jp\leftrightarrow\jm$ and
$\kp\leftrightarrow\km$, and complex-conjugates the coefficients, so
that $e^{\mp2i\omega_0t}$ becomes $e^{\pm2i\omega_0t}$ and the third
term of each bracket is exchanged with the fourth.  Both jump rates and
every observable reported below are invariant under the rotation, so
integrating Eq.~\eqref{eq:dickeH} gives the laboratory-frame answers.

The counter-rotating terms do move the exceptional point, but only just,
and Sec.~\ref{sec:ep} quantifies this: the crossing shifts to
$J=0.9959\,\kappa$ at $\lambda=0.1\,\omega_0$, a displacement of order
$(\lambda/2\omega_0)^2$, because the parametric terms destroy the
conservation of excitation number that makes $\lambda$ drop out of the
Tavis--Cummings branch-point condition.  The charging curves feel the
frequency shift instead, of order $\lambda^2/2\omega_0$, which
is the effect analysed in Sec.~\ref{sec:dickeresults}.

The frame is not a detail.  Removing the free energies while keeping
the counter-rotating terms as they stand strips $\bd\jp$ of its
$2\omega_0$ detuning and turns it into a resonant parametric amplifier
that creates a photon and an atomic excitation together.  For the
isolated battery of Sec.~\ref{sec:part1} at $\bar n_0=4$ and
$\lambda=0.1\,\omega_0$ that reading gives $E_{\max}=8.74\,\omega_0$
and a peak photon number of $12.0$, against $3.90\,\omega_0$ and $4.0$
for the correctly detuned model, which in turn sits $0.36\%$ above the
Tavis--Cummings value $3.886\,\omega_0$.  We integrate
Eq.~\eqref{eq:dickeH} with the phase factors in place, evaluated at the
midpoint of every step.

\subsection{Results}\label{sec:dickeresults}

\begin{figure*}[tbp]
\centering
\includegraphics[width=\textwidth]{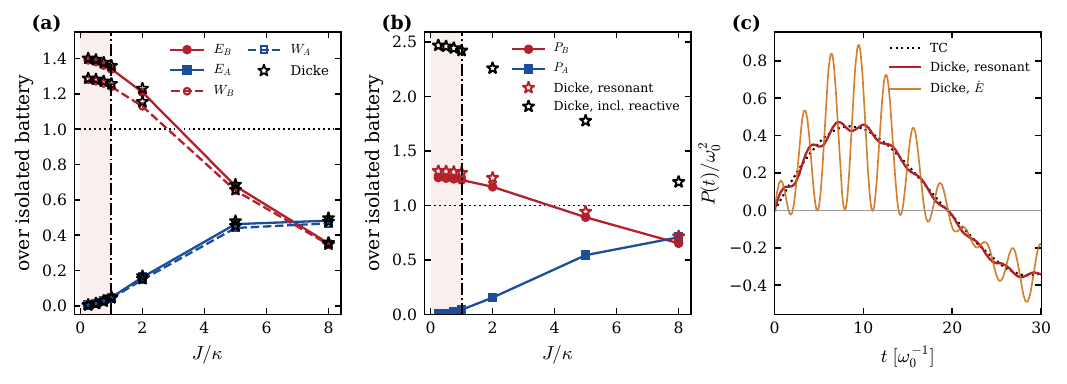}
\caption{(Color online) Dicke batteries against Tavis--Cummings ones,
same parameters as Fig.~\ref{fig:twobat}.  (a) Peak stored energy
(solid, filled markers) and ergotropy (dashed, open markers) of battery
$B$ (red circles) and battery $A$ (blue squares), normalised to the
isolated battery; lines are Tavis--Cummings, stars are Dicke.
(b) Peak charging power: lines are Tavis--Cummings, red stars are the
resonant part $P^{\rm res}_B$ of the Dicke run, black stars the total
flux $P^{\rm tot}_B$.  (c) Charging power of battery $B$ against time
at $J/\kappa=0.25$: Tavis--Cummings (black dotted), the resonant part
of the Dicke run (red solid) and the total Dicke flux (orange solid).
The counter-rotating terms add an oscillation at $2\omega_0$ whose
time average vanishes.}
\label{fig:dicke}
\end{figure*}

Table~\ref{tab:dicke} and Fig.~\ref{fig:dicke} repeat the scan of
Sec.~\ref{sec:twobat} with Eq.~\eqref{eq:dickeH}, at the same
parameters, the same $K=300$ trajectories and the same $301$ time
points; each entry is followed by the Tavis--Cummings value of
Table~\ref{tab:twobat} in parentheses.

\begin{table*}[!tbp]
\caption{Dicke batteries in the same pair.  Peak stored energy $E$,
charging power $P$ and ergotropy $W$ of the gain-side battery $B$ and
the lossy-side battery $A$, from Eq.~\eqref{eq:dickeH}; the
Tavis--Cummings value of Table~\ref{tab:twobat} is given in
parentheses.  $P_B^{\rm res}$ is the resonant part,
Eq.~\eqref{eq:power}, and $P_B^{\rm tot}$ the full energy flux,
Eq.~\eqref{eq:dickepower}; they differ by the reactive term discussed
in the text.  Same parameters as Table~\ref{tab:twobat}, same $K=300$
trajectories, same statistical uncertainty of $\sim2$--$3\%$.}
\label{tab:dicke}
\begin{ruledtabular}
\begin{tabular}{cccccccc}
$J/\kappa$ & $E_B$ & $E_A$ & $P_B^{\rm res}$ & $P_B^{\rm tot}$
 & $W_B$ & $W_A$\\
\colrule
0.25 & $5.44\,(5.41)$ & $0.02\,(0.01)$ & $0.47\,(0.45)$ & $0.89$
 & $4.92\,(4.89)$ & $0.01\,(0.01)$\\
0.5 & $5.41\,(5.37)$ & $0.05\,(0.05)$ & $0.47\,(0.45)$ & $0.88$
 & $4.90\,(4.86)$ & $0.04\,(0.04)$\\
0.75 & $5.35\,(5.29)$ & $0.11\,(0.11)$ & $0.47\,(0.45)$ & $0.88$
 & $4.85\,(4.81)$ & $0.10\,(0.09)$\\
1 & $5.28\,(5.21)$ & $0.19\,(0.18)$ & $0.47\,(0.44)$ & $0.87$
 & $4.80\,(4.74)$ & $0.17\,(0.16)$\\
2 & $4.78\,(4.70)$ & $0.65\,(0.63)$ & $0.45\,(0.42)$ & $0.81$
 & $4.41\,(4.31)$ & $0.58\,(0.56)$\\
5 & $2.66\,(2.63)$ & $1.86\,(1.80)$ & $0.34\,(0.32)$ & $0.64$
 & $2.51\,(2.49)$ & $1.73\,(1.68)$\\
8 & $1.38\,(1.37)$ & $1.93\,(1.87)$ & $0.26\,(0.23)$ & $0.44$
 & $1.33\,(1.32)$ & $1.82\,(1.78)$\\
\end{tabular}
\end{ruledtabular}
\end{table*}

The counter-rotating terms superpose on every
instantaneous quantity a coherent oscillation at $2\omega_0$, of fixed
phase, whose average over a cycle vanishes.  They do not move the
envelope.  On the isolated battery of Sec.~\ref{sec:part1} at
$\bar n_0=4$ this is cleanest: $E_{\rm Dic}(t)-E_{\rm TC}(t)$ is an
oscillation about zero of amplitude $0.183\,\omega_0$, i.e. $4.7\%$ of
the stored energy, whose dominant Fourier component sits at
$1.9\,\omega_0$, while the two peak values differ by only $0.36\%$.  A
number sampled at one instant of an oscillation that averages to zero
over a cycle says little about the envelope, and it is the envelope
that the rest of this paper is about.

In the pair the same holds for the stored energy and the ergotropy.
Across the seven couplings $E_B$ exceeds its Tavis--Cummings value by
$0.6$--$1.7\%$ and $W_B$ by $0.4$--$2.3\%$, while for battery $A$,
once it carries a charge worth speaking of, the excess is $3$--$9\%$
in energy and $2$--$3\%$ in ergotropy.  There is no trend in
$J/\kappa$: the excess is smallest at both ends of the scan and
largest in the middle, which is what one expects from a fixed
excursion sampled at a peak rather than from a change of the envelope.
That it is larger for battery $A$ says only that the same absolute
excursion is a bigger fraction of a smaller charge.  The ordering of
the two batteries, the crossover at which $A$ overtakes $B$ and the
factor of nearly four across the exceptional point are all unchanged,
and so is the fractional gain over the isolated battery: at
$J/\kappa=0.25$ battery $B$ gains $40\%$, $32\%$ and $29\%$ in stored
energy, resonant power and ergotropy, against $39\%$, $26\%$ and
$28\%$ for Tavis--Cummings.  The correction the counter-rotating terms
make to the envelope is of order
$(\lambda/2\omega_0)^2\approx3\times10^{-3}$, and at
$\lambda=0.1\,\omega_0$ that is below what this method can resolve.

The power is where the oscillation becomes visible, and it is worth
stating separately because it changes what the power means.  For the
Dicke battery the energy flux into the atoms is
\begin{equation}
\dot E_B=-\frac{2\lambda\omega_0}{\sqrt N}\,
\mathrm{Im}\mean{\bd\left(\jm-\jp\right)}_{\rm lab},
\label{eq:dickepower}
\end{equation}
of which Eq.~\eqref{eq:power} is only the first term.  The second,
counter-rotating term carries the same oscillation with a much larger
amplitude relative to its own scale: on the isolated battery it
reaches $0.365\,\omega_0^2$ against a resonant peak of
$0.384\,\omega_0^2$, so the peak of $\dot E$ rises from $0.384$ to
$0.707\,\omega_0^2$, very nearly a factor of two, while the stored
energy it produces, the integral of that oscillation, has amplitude
$0.365/2\omega_0=0.18\,\omega_0$ and returns to zero every half
cycle.  The resonant part inherits a smaller share of the same oscillation,
$0.039\,\omega_0^2$ on the isolated battery, which is why its peak
exceeds the Tavis--Cummings one by about seven per cent there and by
$5$--$10\%$ in the pair.  The ratio of the total flux to the resonant
part is $1.70$--$1.89$ over the scan, essentially the factor of two
seen in isolation, and it barely varies with $J/\kappa$, confirming
that the reactive term is a property of the battery--field coupling
alone and not of the environment.

So a Dicke battery in the broken phase stores the same energy, at the
same rate on the cycle average, and delivers the same work as a
Tavis--Cummings one; the counter-rotating terms add a reactive
sloshing of energy between the mode and the atoms at $2\omega_0$ that
carries no net charge.  The practical consequence is that the peak of
the instantaneous power is not a measure of charging performance for a
Dicke battery, and Table~\ref{tab:dicke} therefore lists the resonant
part $P^{\rm res}_B$ of Eq.~\eqref{eq:power} alongside the total flux
$P^{\rm tot}_B$ of Eq.~\eqref{eq:dickepower}.  Fig.~\ref{fig:dicke}(c)
shows both against time at $J/\kappa=0.25$.

\begin{figure*}[tbp]
\centering
\includegraphics[width=\textwidth]{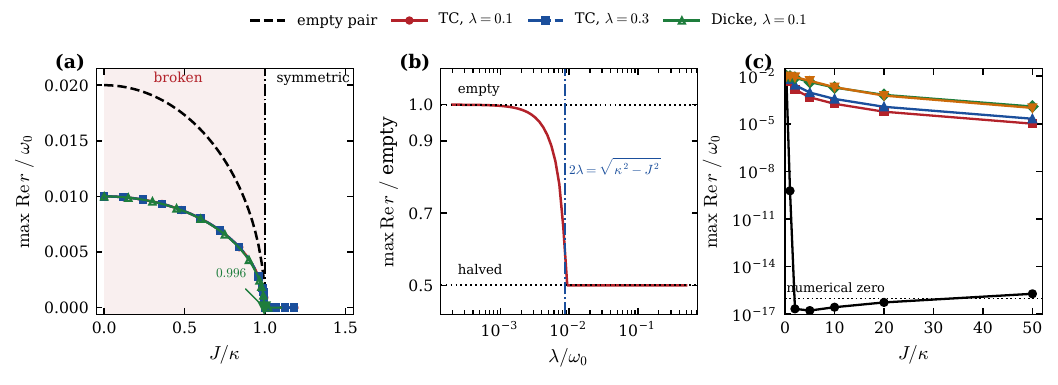}
\caption{(Color online) Linear response of the loaded pair.  (a) Largest
real part of the amplitude eigenvalue against $J/\kappa$: the empty
pair (dashed), the pair loaded with Tavis-Cummings batteries at
$\lambda=0.1\,\omega_0$ (circles) and $0.3\,\omega_0$ (squares), and
the Dicke batteries at $\lambda=0.1\,\omega_0$ from the Floquet
solution of the eight-dimensional Bogoliubov problem (open triangles).
The Tavis-Cummings curves cross zero at $J=\kappa$ and lie a factor of
two below the empty pair throughout the broken phase; the Dicke curve
reaches zero at $0.9959\,\kappa$, marked by the arrow.  (b) The growth
rate at $J/\kappa=0.5$ relative to the empty pair, against
$\lambda$.  Halving sets in where $2\lambda$ overtakes
$\sqrt{\kappa^2-J^2}$.  (c) Unequal loading, $\lambda_A\neq\lambda_B$
with $\lambda_B=0.1\,\omega_0$ fixed: the symmetric phase disappears
entirely, and the residual growth rate decays as a power of $J$ without
ever reaching zero.
\label{fig:ep}}
\end{figure*}

\section{Exceptional point of the loaded pair}\label{sec:ep}

The exceptional point at $J=\kappa$ was located in
Sec.~\ref{sec:part2} from a moment system that closes only because the
cavities are empty.  Put a battery into each of them and the system no
longer closes: the spins are not linear, and $\mean{\hat b\,\jz}$ is
not $\mean{\hat b}\mean{\jz}$.  The effect of loading the cavities on
the exceptional point therefore has to be reopened, and in the limit where
the batteries are barely charged it has an exact answer.  There the
Holstein--Primakoff mapping applies, $\jm\simeq\sqrt N\,\hat c$ and
$\km\simeq\sqrt N\,\hat d$, each battery becomes a bosonic mode, and
neither carries dissipation of its own.  The pair is then a four-mode
linear network whose eigenvalue problem can be solved in closed form.

\subsection{Linearised model}

Take the four mode amplitudes as a column
$\mathbf v=(\mean{\hat a},\mean{\hat b},\mean{\hat c},\mean{\hat d})$,
with $\hat c$ the battery in the gain cavity and $\hat d$ the one in
the lossy cavity.  The first two rows follow from operator identities
and hold at any excitation: the adjoint dissipators give
$\mathcal D^{\dagger}[\hat a]\,\hat a=-\tfrac{\gamma}{2}\hat a$ and
$\mathcal D^{\dagger}[\bd]\,\hat b=+\tfrac{\gamma}{2}\hat b$, and it
is the second of these, not the effective Hamiltonian, that carries the
amplification.  The last two rows are where the linearisation sits,
$\mean{\hat b\,\jz}\simeq-\tfrac N2\mean{\hat b}$.  The four equations
are
\begin{align}
\dot a&=-\kappa\,a-iJb-i\lambda d,&
\dot b&=+\kappa\,b-iJa-i\lambda c,\nonumber\\
\dot c&=-i\lambda b,& \dot d&=-i\lambda a.
\label{eq:drift}
\end{align}

\subsection{Location of the branch point}

Eliminating $c$ and $d$ and writing $u\equiv r+\lambda^2/r$ reduces
the first two rows to $(u+\kappa)a=-iJb$ and $(u-\kappa)b=-iJa$, so
that
\begin{equation}
u^2=\kappa^2-J^2,\qquad
r=\tfrac12\left(u\pm\sqrt{u^2-4\lambda^2}\right),
\label{eq:branch}
\end{equation}
the second relation being $r^2-ur+\lambda^2=0$.  The coupling
$\lambda$ has dropped out of the first relation altogether, and with
it every statement about the batteries.  Their loading is equal and
they carry no dissipation, so they cannot alter the balance of gain
against loss, and the branch point $u_+=u_-$ stays at $J=\kappa$.
Table~\ref{tab:ep} confirms this over two decades of $\lambda$: the
location never moves by more than $8\times10^{-6}$, which is the
accuracy of the eigensolver at a degenerate point rather than a
physical displacement.

\begin{table}[!tbp]
\caption{Exceptional point of the loaded pair.  For each coupling, the
location at which $\max\,\mathrm{Re}\,r$ vanishes and the four
eigenvalues there, which are doubly degenerate and fourfold degenerate
for the empty pair.  The last column is the growth rate at
$J/\kappa=0.5$ in units of $\omega_0$; every loaded entry is exactly
one half of the empty-pair value.  The closed form of
Eq.~\eqref{eq:branch} reproduces each number to machine precision, the
largest deviation over eighteen couplings and eigenvalues being
$2\times10^{-16}$.}
\label{tab:ep}
\begin{ruledtabular}
\begin{tabular}{cccc}
$\lambda/\omega_0$ & EP at $J/\kappa$ & $r$ at the EP &
$\max\,\mathrm{Re}\,r$\\
\colrule
$0$    & $1.0000000$ & $0$          & $0.017321$\\
$0.01$ & $1.0000000$ & $\pm0.010i$ & $0.008660$\\
$0.05$ & $1.0000000$ & $\pm0.050i$ & $0.008660$\\
$0.1$  & $1.0000001$ & $\pm0.100i$ & $0.008660$\\
$0.3$  & $1.0000003$ & $\pm0.300i$ & $0.008660$\\
$1.0$  & $1.0000080$ & $\pm1.000i$ & $0.008660$\\
\end{tabular}
\end{ruledtabular}
\end{table}

The spectrum is not the spectrum of the empty pair for all that, and
two features of it are worth recording.  First, at $J=\kappa$ the
branch point has $u=0$ and so $r=\pm i\lambda$: the exceptional point
no longer sits at zero frequency but at the polariton frequency, and
what vanishes there is the real part rather than the eigenvalue.
Table~\ref{tab:ep} finds $\pm0.01\,i$, $\pm0.05\,i$, $\pm0.1\,i$,
$\pm0.3\,i$ and $\pm1.0\,i$ at the five couplings.  Second, the growth
rate is halved throughout the broken phase.  This happens whenever
$u^2<4\lambda^2$, that is whenever
$2\lambda>\sqrt{\kappa^2-J^2}$, because then the square root in
Eq.~\eqref{eq:branch} is imaginary and $\max\,\mathrm{Re}\,r=\tfrac12u$.
Our parameters satisfy the condition across the whole broken phase,
$\lambda=0.1\,\omega_0$ exceeding $\kappa=0.02\,\omega_0$ fivefold, and
Fig.~\ref{fig:ep}(b) locates the crossover at
$\lambda=\tfrac12\sqrt{\kappa^2-J^2}=0.0087\,\omega_0$ for
$J/\kappa=0.5$.  The reason is that each cavity mode hybridises with a
lossless atomic mode, so that only half of every polariton is damped
or amplified.  The square root has a branch point of its own at
$\kappa^2-J^2=4\lambda^2$, but it exists only for
$\lambda<\kappa/2$ and does not bound the broken phase, growth
persisting up to $J=\kappa$ either way.

\subsection{Dicke batteries in the same network}

The counter-rotating terms add to Eq.~\eqref{eq:drift} the parametric
pieces $\lambda e^{2i\omega_0t}\bd\hat c^{\dagger}$ and
$\lambda e^{-2i\omega_0t}\hat b\,\hat c$, and the matching pair for
$\hat a$ and $\hat d$.  Detuned by $2\omega_0$ they do not conserve the
excitation number, and that is what spoils the cancellation: with a
term that creates two excitations at once the battery is no longer a
passive lossless load, and $\lambda$ no longer drops out.  The problem
is periodic, so we solve it as a Floquet problem on the
eight-dimensional Bogoliubov system
$(\hat a,\hat b,\hat c,\hat d,\ad,\bd,\hat c^{\dagger},\hat
d^{\dagger})$ over one period $\pi/\omega_0$.

The displacement is small and of the expected order.
Table~\ref{tab:epdicke} puts the crossing at $0.9998\,\kappa$,
$0.9959\,\kappa$ and $0.9730\,\kappa$ for $\lambda=0.02$, $0.1$ and
$0.3\,\omega_0$, and in the weak-coupling limit the displacement
approaches $-2(\lambda/2\omega_0)^2$, the higher-order corrections
being visible at the largest couplings.  Away from the crossing the
real parts are those of the Tavis-Cummings problem to within
$2.7\times10^{-6}\,\omega_0$ at $J/\kappa=0.25$ and
$1.2\times10^{-5}\,\omega_0$ at $J/\kappa=0.5$, rising to
$3\times10^{-4}\,\omega_0$ only where the Tavis-Cummings rate is itself
falling to zero.  The exponents at the crossing sit at
$\pm0.0945\,i$ and $\pm0.1067\,i$ against the Tavis-Cummings
$\pm0.1\,i$, a splitting of $\pm\lambda^2/2\omega_0$ of the
Bloch--Siegert type that moves frequencies alone.

\begin{table}[!tbp]
\caption{Exceptional point with Dicke batteries, from the Floquet
solution of the eight-dimensional Bogoliubov problem over one period
$\pi/\omega_0$.  The Tavis-Cummings crossing is at $J/\kappa=1$ for
every coupling in the table.}
\label{tab:epdicke}
\begin{ruledtabular}
\begin{tabular}{cccc}
$\lambda/\omega_0$ & EP at $J/\kappa$ & displacement &
displacement$/({\lambda}/{2\omega_0})^2$\\
\colrule
$0.02$ & $0.999808$ & $-1.92\times10^{-4}$ & $-1.92$\\
$0.05$ & $0.998865$ & $-1.13\times10^{-3}$ & $-1.82$\\
$0.10$ & $0.995859$ & $-4.14\times10^{-3}$ & $-1.66$\\
$0.15$ & $0.991456$ & $-8.54\times10^{-3}$ & $-1.52$\\
$0.20$ & $0.986012$ & $-1.40\times10^{-2}$ & $-1.40$\\
$0.30$ & $0.973006$ & $-2.70\times10^{-2}$ & $-1.20$\\
\end{tabular}
\end{ruledtabular}
\end{table}

\subsection{The requirement of identical batteries}

The cancellation in Eq.~\eqref{eq:branch} also requires the two
batteries to be identical, and this is not a technical convenience.
Keeping the couplings distinct, $\lambda_B$ in the gain cavity and
$\lambda_A$ in the lossy one, the same elimination gives
$(r^2+\lambda_A^2)(r^2+\lambda_B^2)+\kappa r(\lambda_B^2-\lambda_A^2)
+(J^2-\kappa^2)r^2=0$, of which Eq.~\eqref{eq:branch} is the
$\lambda_A=\lambda_B$ case.  The term linear in $r$ appears whenever
$\lambda_A\neq\lambda_B$, and it removes the symmetry under
$r\to-r$ that lets all four roots have vanishing real part.  No
symmetric phase survives at any $J$: for $\lambda_A=0.105\,\omega_0$
against $\lambda_B=0.1\,\omega_0$ the largest real part is
$4.6\times10^{-4}\,\omega_0$ at $J/\kappa=5$ and still
$1.0\times10^{-5}\,\omega_0$ at $J/\kappa=50$, against
$10^{-17}\,\omega_0$, the level of numerical zero, for the symmetric
loading.  Fig.~\ref{fig:ep}(c) shows the family.

The extreme case $\lambda_A=0$, a battery in the gain cavity and none
in the lossy one, can be solved outright: one root is $r=0$ and the
other three obey
$r^3+(\lambda_B^2+J^2-\kappa^2)\,r+\kappa\lambda_B^2=0$, whose real
root is negative for every $J>\kappa$, so the complex pair has a
positive real part, asymptotically $\kappa\lambda_B^2/2J^2$.  At
$J/\kappa=50$ that predicts $1.0\times10^{-4}\,\omega_0$ against the
measured $9.9\times10^{-5}\,\omega_0$.  It is worth noting what this
says about the arrangement we discarded: a single battery breaks the
balance by its mere presence, whatever the coupling, and the
directionality of the broken phase then has no phase transition to
answer to.

\section{Numerical method}\label{sec:method}

We integrate the master equation with the Monte-Carlo wave-function
method~\cite{Dalibard1992,Plenio1998} in a batched implementation:
$K$ trajectories are stored as the columns of a dense $(D,K)$ array and
propagated simultaneously with a fourth-order Taylor expansion of
$\exp(\hat H_{\rm eff}\Delta t)$, with $\hat H_{\rm eff}=\hat H-\frac{i}{2}\sum_l
\hat L_l^\dagger \hat L_l$.  The time step is adapted so that the largest
jump rate times $\Delta t$ stays below $0.03$.  We use $K=300$
trajectories, a Fock cutoff of $40$ photons in cavity $\hat b$ and $8$ in
cavity $\hat a$, and $301$ time points.  For the two-battery model of
Eq.~\eqref{eq:symH} the basis is
$\ket{n_a,n_b,m_J,m_K}$ and the Hilbert space dimension is
$D=9\times41\times11\times11=44649$.  The reduced state of either battery
is obtained by tracing out the other three factors of every trajectory.

Three checks bound the error.  First, the equal-time
power~\eqref{eq:power} reproduces
$\mathrm{d}E/\mathrm{d}t$ of the isolated battery to
$4\times10^{-3}\,\omega_0^2$, which is the discretisation error of the
finite-difference reference.  Second, the population
of the top Fock layer of cavity $\hat b$, monitored by a dedicated
projector, never exceeds $1.9\times10^{-10}$ anywhere in the $J$ scan.
Third, the statistics:
repeating the $J/\kappa=0.5$ case with $K=900$ changes
$E_B$, $P_B$ and $W_B$ by $0.2\%$, $1.8\%$ and $0.3\%$;
we therefore quote $K=300$ results with a $\sim2$--$3\%$ statistical
uncertainty, larger only where the charge itself is small, as for
battery $A$ in the broken phase.  Appendix~\ref{app:numeric} gives the
details.

\section{Conclusions}\label{sec:concl}

We have studied the charging of Dicke and Tavis--Cummings quantum
batteries from the collective ground state, first in isolation and then
coupled to a $\mathcal{PT}$-symmetric lossy--gain pair, with an
identical battery in each cavity to keep the symmetry intact.

For the isolated battery we proved the short-time laws
$\mean{\Ne(t)}=\lambda^2\bar n_0t^2$ (TC) and
$\lambda^2(4\bar n_0+1)t^2$ (Dicke), recovered the numerical
coefficients to $10^{-5}$ (TC) and $0.15\%$ (Dicke), and established the
exact bound
$E_{\max}\le\omega_0\min(\nb,N)$ for the TC model.  Numerically the
bound is saturated to $80$--$100\%$ for $\nb\in[1,50]$; the Dicke model
at $\lambda=0.3\,\omega_0$ violates it by up to $3.9\%$, a direct
signature of counter-rotating ``virtual'' photons.  The practical
conclusion of Sec.~\ref{sec:part1} is that a larger initial reservoir both speeds up
and strengthens the charge, but only up to the ceiling
$\omega_0\min(\nb,N)$: once the reservoir is exhausted nothing further
happens.

For the $\mathcal{PT}$-symmetric pair we derived the exact closed
moment system and its characteristic polynomial
$-r(r^2-\gamma^2+4J^2)$, locating the exceptional point at
$J=\kappa=\gamma/2$, the broken-phase growth rate
$2\sqrt{\kappa^2-J^2}$ and the unbroken-phase secular drift
$\gamma J^2/[2(J^2-\kappa^2)]$.  Every one of these expressions was
verified against an independent numerical construction to
$10^{-13}$ or better.

For the batteries in the pair we restored the strict
$\mathcal{PT}$ symmetry by putting an identical battery in each cavity.
The broken phase delivers essentially all the energy to the gain-side
battery, the lossy-side one staying empty, while in the unbroken phase
the two batteries split a smaller total and the lossy-side one overtakes
the gain-side one: the directionality comes from the initial state, not
from any asymmetry of the Liouvillian.  The exceptional point separates
two charging regimes that differ by a factor of order $4$ in stored
energy and ergotropy.  Relative to an isolated battery, the broken phase
adds
$39\%$ energy, $26\%$ power and $28\%$ ergotropy to the gain-side
battery, while the deep unbroken phase removes
$65\%$, $35\%$ and $65\%$ and hands what is left to the lossy-side one.
The mechanism is the
competition between photon accumulation in the gain cavity and photon
drainage into the lossy cavity, quantified through
$\max_t\mean{\hat n_a}$ and the total excitation number; the current that
drains cavity $\hat b$ is what feeds battery $A$.

Replacing the Tavis--Cummings coupling by the Dicke one, that is,
restoring the counter-rotating terms, does not change the charging:
the stored energy and the ergotropy of both batteries agree with the
Tavis--Cummings values to within the statistical error across the
whole scan, and the correction to the envelope is of order
$(\lambda/2\omega_0)^2\approx3\times10^{-3}$.  The counter-rotating
terms add a coherent oscillation at $2\omega_0$
whose average over a cycle vanishes; it is worth a few per cent at the
peak of the stored energy and it nearly doubles the peak of $\dot E$,
because there it is as large as the resonant flux itself.  The two
models can be treated together at this coupling, and the peak of the
instantaneous power is not a quantity with which to compare them.

We also took up the exceptional point itself and resolved it in the
low-excitation limit, where the problem reduces to a
four-mode linear network that can be solved in closed form.  The
Tavis--Cummings batteries leave the crossing at $J=\kappa$ for every
coupling, because they load the two cavities equally and carry no
dissipation, so the balance of gain against loss is untouched; the
spectrum is renormalised all the same, the critical eigenvalues moving
to $\pm i\lambda$ and the broken-phase growth rate halving, since each
cavity mode now hybridises with a lossless atomic mode.  The Dicke
batteries do move the crossing, but only by an amount of order
$(\lambda/2\omega_0)^2$, four tenths of a per cent at the coupling used
here, because the counter-rotating terms destroy the conservation of
excitation number on which the cancellation rests.  Unequal loading is
fatal to the symmetric phase at any $J$, and a battery in the gain
cavity alone is already enough to produce it.

We also measured how much of the stored energy is actually recoverable.
The ergotropy of the isolated battery stays within a few per cent of the
stored energy, with an extractable fraction above $94\%$ for all
reservoirs except the one that just suffices to invert the battery,
$\bar n_0\simeq N$, where it falls to $91\%$.  Coupling the batteries to
the pair lowers the fraction of the gain-side battery to $90\%$ in the
broken phase and raises it back to $96\%$ deep in the unbroken phase,
while the lossy-side battery tracks it about half a per cent below; the
extra energy supplied by the gain is partly paid for in entropy, and
roughly seven tenths of the added stored energy is nonetheless recovered
as added work.  Using the convexity of the ergotropy we separated the
locked energy into an entanglement-locked part, at most about three per
cent of the stored energy, and a classically locked part due to the
unread jump record, which is about twice as large; the same holds for
the lossy-side battery where it charges.  A
$\mathcal{PT}$-broken environment makes a good charger, if a slightly
dirty one: it delivers more work, not merely more energy,
and the part it fails to deliver is mostly classical disorder rather
than entanglement.

Three extensions suggest themselves.  First, a semianalytic treatment of
the composite problem, combining the moment system with a
Holstein--Primakoff description of the battery, would make the
$\sim\sqrt{\nb}$ scaling of the power predictive rather than
descriptive.  Second, the Dicke
comparison of Sec.~\ref{sec:dicke} was made at one coupling; the
counter-rotating terms violate the bound~\eqref{eq:bound} in isolation
only at ultrastrong coupling, and it is there that they should also
move the two-battery charge, a regime whose photon cutoff is larger
than the one used here.  Third, a systematic
optimisation over $J$, $\gamma$ and $\lambda$ in the presence of the EP
enhancement of the susceptibility is a natural next step.

\begin{acknowledgments}
This work was supported by NSFC (Grant No.~12205092), the Hunan
Provincial Natural Science Foundation of China (Grant
Nos.~2026JJ50100 and 2026JJ50513) and the Scientific Research Fund of
Hunan Provincial Education Department (Grant No.~24B0670).
\end{acknowledgments}

\appendix

\section{Short-time expansion}\label{app:shorttime}

Let $\hat A$ be Hermitian, $\ket{\psi_0}$ the initial state and
$\hat U(t)=e^{-i\hat H t}$ the propagator.  Expanding
$\mean{\hat A(t)}=\bra{\psi_0}\hat U^\dagger(t)\hat A\hat U(t)\ket{\psi_0}$ to second order,
\begin{align}
\mean{\hat A(t)}&=\mean{\hat A}
+i\,t\,\mean{[\hat H,\hat A]}\nonumber\\
&\quad+t^2\left(\mean{\hat H\hat A\hat H}-\tfrac12\mean{\hat A\hat H^2+\hat H^2\hat A}\right)+O(t^3).
\label{eq:appexpansion}
\end{align}
Take $\hat A=\Ne=\jz+j$ and $\ket{\psi_0}=\ket{\mathrm{field}}
\otimes\ket{j,-j}$.  Then $\Ne\ket{\psi_0}=0$, hence
$\mean{\Ne}=0$, $\mean{[\hat H,\Ne]}=\mean{\hat H\Ne}-\mean{\Ne \hat H}=0$ and
$\mean{\Ne \hat H^2}=\mean{\hat H^2\Ne}=0$, because
$\bra{\psi_0}\Ne=0$.  Only the $\mean{\hat H\Ne \hat H}$ term survives:
\begin{equation}
\mean{\Ne(t)}=t^2\bra{\psi_0}\hat H\Ne \hat H\ket{\psi_0}+O(t^4).
\end{equation}

The free terms of Eqs.~\eqref{eq:htc} and~\eqref{eq:hdic} drop out of
$\mean{\hat H\Ne\hat H}$: acting on $\ket{\psi_0}$ they return either
$\ket{\psi_0}$ or a state with the atoms still in $\ket{j,-j}$, and
$\Ne$ annihilates both.  Only the coupling part remains, in the
rotating frame.

For the TC model, with
$\hat H_{\rm TC}=(\lambda/\sqrt N)(\hat a\jp+\ad\jm)$ and
$\jp\ket{j,-j}=\sqrt N\ket{j,-j+1}$,
\begin{equation}
\hat H_{\rm TC}\ket{\psi_0}=\lambda\,\hat a\ket{\rm field}\otimes\ket{j,-j+1},
\end{equation}
and $\Ne\ket{j,-j+1}=\ket{j,-j+1}$.  Therefore
\begin{equation}
\bra{\psi_0}\hat H\Ne \hat H\ket{\psi_0}
=\lambda^2\bra{\rm field}\ad \hat a\ket{\rm field}
=\lambda^2\bar n_0,
\end{equation}
which is Eq.~\eqref{eq:shortTC}.

For the Dicke model, with
$\hat H_{\rm Dic}=(\lambda/\sqrt N)(\hat a+\ad)(\jp+\jm)$,
\begin{equation}
\hat H_{\rm Dic}\ket{\psi_0}
=\lambda\,(\hat a+\ad)\ket{\rm field}\otimes\ket{j,-j+1},
\end{equation}
so that
\begin{equation}
\bra{\psi_0}\hat H\Ne \hat H\ket{\psi_0}
=\lambda^2\mean{(\hat a+\ad)^2}_0,
\end{equation}
which is Eq.~\eqref{eq:shortDicke}.  For a coherent state with real
$\alpha$, $\mean{(\hat a+\ad)^2}_0=\alpha^2+\alpha^{*2}+2\bar n_0+1
=4\bar n_0+1$; for a Fock state $\ket n$ it equals $2n+1$.

The semiclassical limit is obtained by replacing $\hat a$ by $\sqrt{\nb}$ in
Eq.~\eqref{eq:htc} gives
$\hat H_{\rm sc}=(\lambda\sqrt{\nb}/\sqrt N)(\jp+\jm)
=2\lambda\sqrt{\nb/N}\,\hat J_x$, a rotation of the collective spin about
$x$ at $\Omega=2\lambda\sqrt{\nb/N}$.  Starting from
$\ket{j,-j}$,
$\mean{\Ne(t)}=\frac N2[1-\cos(\Omega t)]$, whence
Eq.~\eqref{eq:semi}.

\section{The moment system of the $\mathcal{PT}$ pair}\label{app:moments}

We use the adjoint master equation
$\frac{d}{dt}\mean{\hat O}
=i\mean{[\hat H,\hat O]}+\sum_l\mean{\hat L_l^\dagger \hat O\hat L_l-\tfrac12\{\hat L_l^\dagger \hat L_l,\hat O\}}$
with $\hat H=J(\ad \hat b+\hat a\bd)$, $\hat L_1=\sqrt\gamma\,\hat a$, $\hat L_2=\sqrt\gamma\,\bd$.
The following elementary commutators are used throughout:
\begin{equation}
[\ad \hat a,\hat a]=-\hat a,\quad [\ad \hat a,\hat a^\dagger]=\hat a^\dagger,\quad
[\jz,\jp]=\jp,
\end{equation}
and, with $\hat X=\ad \hat b$ and $\hat X^\dagger=\hat a\bd$,
\begin{equation}
[\hat X^\dagger,\hat X]=N_b-N_a,\qquad [\hat H,\hat X]=J(N_b-N_a).
\end{equation}

The Hamiltonian parts follow from $[\hat H,\hat X]=J(N_b-N_a)$ and
$[\hat H,\hat X^\dagger]=-J(N_b-N_a)$,
\begin{equation}
i[\hat H,\hat X-\hat X^\dagger]=2iJ(N_b-N_a),
\end{equation}
which gives the Hamiltonian parts
$-JS$ of $\dot N_a$, $+JS$ of $\dot N_b$ and $-2J(N_b-N_a)$ of
$\dot S$ with $S=i\mean{\hat X-\hat X^\dagger}$.

Only the loss channel contributes to $\dot N_a$:
$\mathcal{D}^\dagger[\sqrt\gamma\,\hat a][\ad \hat a]
=\gamma\bigl(\ad\ad\hat a\hat a-(\ad \hat a)^2\bigr)$; since
$\ad\ad\hat a\hat a=(\ad \hat a)^2-\ad \hat a$, this is $-\gamma N_a$.

Only the gain channel contributes to $\dot N_b$:
$\mathcal{D}^\dagger[\sqrt\gamma\,\bd][\bd \hat b]
=\gamma\,\hat b\,\bd \hat b\,\bd-\tfrac\gamma2\{\hat b\bd,\bd \hat b\}
=\gamma(\bd \hat b+1)$.

The two channels act on $\hat X=\ad \hat b$ with opposite signs and
cancel exactly, which is why $S$ has no damping term.  For the loss
channel,
\begin{align}
\mathcal{D}^\dagger[\sqrt\gamma\,\hat a]\hat X
&=\gamma\bigl[\ad \hat X\hat a-\tfrac12\{\ad \hat a,\hat X\}\bigr]\nonumber\\
&=\gamma\bigl[\ad\ad\hat a\,\hat b-\ad(\ad \hat a+\tfrac12)\hat b\bigr]
=-\tfrac\gamma2\ad \hat b,\label{eq:dissXa}\\
\mathcal{D}^\dagger[\sqrt\gamma\,\bd]\hat X
&=\gamma\bigl[\hat b\hat X\bd-\tfrac12\{\hat b\bd,\hat X\}\bigr]\nonumber\\
&=\gamma\bigl[\ad(\bd \hat b+2)\hat b-\ad(\bd \hat b+\tfrac32)\hat b\bigr]
=+\tfrac\gamma2\ad \hat b,\label{eq:dissXb}
\end{align}

Collecting,
\begin{equation}
\frac{d}{dt}\begin{pmatrix}N_a\\ N_b\\ S\end{pmatrix}
=\begin{pmatrix}-\gamma&0&-J\\ 0&+\gamma&+J\\ 2J&-2J&0\end{pmatrix}
\begin{pmatrix}N_a\\ N_b\\ S\end{pmatrix}
+\begin{pmatrix}0\\ \gamma\\ 0\end{pmatrix}.
\label{eq:appmatrix}
\end{equation}
The characteristic polynomial of the matrix in Eq.~\eqref{eq:appmatrix}
is $-r(r^2-\gamma^2+4J^2)$, giving
$r=0$ and $r_\pm=\pm2\sqrt{\kappa^2-J^2}$ with
$\kappa=\gamma/2$.

The right and left null vectors of the generator
are $w=(1,1,-\gamma/J)^{\mathsf T}$ and
$u=(1,1,\gamma/(2J))^{\mathsf T}$.  Projecting the constant vector
$(0,\gamma,0)^{\mathsf T}$ onto $w$ along the remaining eigenmodes
gives the linear-in-$t$ drift
\begin{equation}
\frac{\mathrm{d}}{\mathrm{d}t}\bigl(N_a,N_b\bigr)
\;\longrightarrow\;
\left(\frac{u^{\mathsf T}c}{u^{\mathsf T}w},\,
\frac{u^{\mathsf T}c}{u^{\mathsf T}w}\right)
=\frac{\gamma J^2}{2(J^2-\kappa^2)}\,(1,1),
\end{equation}
i.e. Eq.~\eqref{eq:secular}.

For the numerical verification, Eq.~\eqref{eq:appmatrix} was
integrated with matrix exponentials and compared, over $t\le50$, with
the trajectory obtained from an independently constructed
$15$-dimensional moment generator (least-squares fit of the adjoint
Liouvillian on a truncated Fock space with $n_{\max}=90$).  For
coherent, vacuum and Fock initial states and for
$J/\kappa=0.25,0.5,0.9,1,1.5,3,5$ the maximum absolute discrepancy in
$N_a$, $N_b$ and $S$ never exceeded $2.5\times10^{-12}$.  The secular
slope of Eq.~\eqref{eq:secular} was reproduced by a linear fit of the
long-time solution to $0.1\%$, and the broken-phase rate
$2\sqrt{\kappa^2-J^2}$ to machine precision.

\section{Numerical convergence}\label{app:numeric}

The composite Hilbert space is
$\mathcal{H}_a\otimes\mathcal{H}_b\otimes\mathcal{H}_{J}\otimes
\mathcal{H}_{K}$ with
$\dim\mathcal{H}_a=9$, $\dim\mathcal{H}_b=41$ and
$\dim\mathcal{H}_{J}=\dim\mathcal{H}_{K}=11$, a total of
$D=44649$.  We monitor the population
$\mean{\hat P_{\rm top}}$ of the highest Fock layer of cavity $\hat b$ with the
projector $\hat P_{\rm top}=\mathbb{I}_a\otimes\ket{n_{\max}}\bra{n_{\max}}
\otimes\mathbb{I}_J\otimes\mathbb{I}_K$.  Over all seven cases of the
$J$ scan the maximum
of $\mean{\hat P_{\rm top}}$ was $1.9\times10^{-10}$, so the truncation is
never active.  A run with the cutoff lowered to $20$ photons was rejected:
it raises the top-layer population to $2\times10^{-2}$.

The MCWF engine adapts $\Delta t$ so that
$\max_k\gamma\mean{\hat n_a+\hat n_b+1}_k\Delta t\le0.03$, with a floor
$\Delta t\ge10^{-4}\,\omega_0^{-1}$ and a cap $\Delta t\le0.05$.
Halving the cap changes $E_{\max}$ by less than the statistical error;
doubling the initial step to $0.1\,\omega_0^{-1}$ shifts $E_B$ by
$2.7\%$, and lowering the photon cutoff to $20$ raises the top-layer
population to $2\times10^{-2}$, so both conservative settings are kept.

The Dicke runs of Sec.~\ref{sec:dicke} integrate a Hamiltonian that
depends on time through the phases of Eq.~\eqref{eq:dickeH}, evaluated
at the midpoint of each step.  Their cap is
$\Delta t\le0.04\,\omega_0^{-1}$.  With the jumps switched off, where
the propagation is deterministic and free of Monte-Carlo noise,
halving the cap from $0.04$ to $0.02$ moves $E_B$ by
$1\times10^{-4}\,\omega_0$ and halving it again by
$4\times10^{-6}\,\omega_0$, so the step size contributes far less than
the statistical error.  Over their seven cases the top-layer
population never exceeds $6\times10^{-9}$.

Table~\ref{tab:convergence} compares
the $J/\kappa=0.5$ case at $K=300$ and $K=900$ trajectories.

\begin{table}[b]
\caption{MCWF convergence of the two-battery model at $J/\kappa=0.5$,
peak values of the gain-side battery $B$.}
\label{tab:convergence}
\begin{ruledtabular}
\begin{tabular}{lccc}
 & $E_B/\omega_0$ & $P_B/\omega_0^2$ & $W_B/\omega_0$\\
\colrule
$K=300$ & $5.370$ & $0.4484$ & $4.855$\\
$K=900$ & $5.361$ & $0.4567$ & $4.840$\\
deviation & $0.2\%$ & $1.8\%$ & $0.3\%$\\
\end{tabular}
\end{ruledtabular}
\end{table}

Finally, we check the consistency of the observables.  For both
batteries in every case the pointwise
inequality $W(t)\le E(t)$ holds to machine precision, as
Eq.~\eqref{eq:ergotropy} requires, and the isolated battery satisfies
$E_{\max}=3.886\le\omega_0\min(\nb,N)=4$ in accordance with
Eq.~\eqref{eq:bound}.  We also verified the frame change of
Sec.~\ref{sec:models} directly: propagating
Eq.~\eqref{eq:symH} and its laboratory-frame counterpart, free
energies included, over the full space for
$J/\kappa=0.25$, $1$ and $8$ gives the same
$\mean{\hat n_a}$, $\mean{\hat n_b}$ and stored energies to
$10^{-8}$, which is the truncation level of the propagator and not a
physical difference.

\bibliography{refs}

\end{document}